\documentclass[12pt,aps,preprint,nofootinbib]{revtex4}

\usepackage{graphicx}
\usepackage{cancel}
\usepackage{amssymb}
\usepackage{textcomp}
\usepackage{amsmath}
\usepackage{bm}
\usepackage{times}
\usepackage{epsfig}
\usepackage{epstopdf}
\usepackage{color}
\usepackage{multirow}
\usepackage[colorlinks=true, pdfstartview=FitV, linkcolor=blue, citecolor=blue, urlcolor=blue]{hyperref}

\def\lsim{\mathrel{\raise.3ex\hbox{$<$\kern-.75em\lower1ex\hbox{$\sim$}}}}
\def\gsim{\mathrel{\raise.3ex\hbox{$>$\kern-.75em\lower1ex\hbox{$\sim$}}}}

\def\beq{\begin{equation}}
\def\eeq{\end{equation}}
\def\be{\begin{equation}}
\def\ee{\end{equation}}
\def\bea{\begin{eqnarray}}
\def\eea{\end{eqnarray}}

\def\h{h^0}
\def\H{H^0}
\def\Hpm{H^\pm}
\def\A{A^0}
\def\mh{m_{h^0}}
\def\mH{m_{H^0}}
\def\mHpm{m_{H^\pm}}
\def\ma{m_{A}}

\def\gev{\,{\rm GeV}}

\def\to{\rightarrow}

\def\ww{W^{+}W^{-}}
\def\tautau{\tau^{+}\tau^{-}}

\begin{document}

\preprint{~~PITT-PACC-1205}

\title{Pair Production of MSSM Higgs Bosons\\ in the Non-decoupling Region at the LHC}


\author{Neil D. Christensen$^{\bf a}$}

\author{Tao Han$^{\bf a}$}

\author{Tong Li$^{\bf b,c}$}

\affiliation{
$^{\bf a}$  Pittsburgh Particle physics, Astrophysics, and Cosmology Center, Department of Physics and Astronomy, University of Pittsburgh, 3941 O'Hara St., Pittsburgh, PA 15260, USA\\
$^{\bf b}$  Bartol Research Institute, Department of Physics and Astronomy,
University of Delaware, Newark, DE 19716, USA\\
$^{\bf c}$  ARC Centre of Excellence for Particle Physics at the Terascale, School of Physics, Monash University, Melbourne, Victoria 3800, Australia
}

\begin{abstract}
We consider the Higgs boson signals from pair production at the LHC within the framework of the MSSM in the  non-decoupling (low-$\ma$) region. In light of the recent observation of a SM-like Higgs boson, we argue that the exploration for Higgs pair production at the LHC is a crucial next step to probe the MSSM Higgs sector. We emphasize that the production of $H^\pm \A $ and $H^{+}H^{-}$ depends only on the electroweak gauge couplings while all the leading Higgs production channels via gluon fusion, vector-boson fusion, and Higgsstrahlung depend on additional free Higgs sector parameters. In the non-decoupling region, the five MSSM Higgs bosons are all relatively light and pair production signals may be accessible.
We find that at the 8 TeV LHC, a $5\sigma$ signal for
$\Hpm \A,\ \Hpm \h \to \tau^{\pm}\nu\ b\bar b$ and $H^{+}H^{-}\to \tau^{+}\nu\  \tau^{-}\nu$ are achievable with an integrated luminosity of 7 (11) fb$^{-1}$ and 24 (48) fb$^{-1}$, respectively for $m_A=95\ (130)$ GeV. At the 14 TeV LHC, a $5\sigma$ signal for these two channels would require as little as 4 (7) fb$^{-1}$ and 10 (19) fb$^{-1}$, respectively.

\end{abstract}

\maketitle

\section{Introduction}

%
%
Recently, both the ATLAS Collaboration \cite{ATLASHnew} and the CMS Collaboration \cite{CMSHnew} at the LHC experiments have reported the observation of a new bosonic particle consistent with the Standard Model (SM) Higgs boson at a mass
\begin{eqnarray}
&& 126.0 \pm 0.4\ ({\rm stat.}) \pm 0.4\ ({\rm syst.})\ {\rm GeV}\quad ({\rm ATLAS}), \\
&& 125.3 \pm 0.4\ ({\rm stat.}) \pm 0.5\ ({\rm syst.})\ {\rm GeV}\quad ({\rm CMS}),
\end{eqnarray}
with a local significance of $5.9\sigma$ and $5.0\sigma$, respectively. The current signal sensitivity is largely due to the Higgs decay channels to $\gamma\gamma,\ ZZ$ and to a lesser extent $WW$, while the signals of the fermionic channels $\tau^+\tau^-,\  b\bar b$ are still very weak. There is thus a hope that more detailed studies of the Higgs boson properties would open a window to new physics associated with the electroweak symmetry breaking sector.

Within the framework of the Minimal Supersymmetric Standard Model (MSSM) \cite{Gunion:1989we,Djouadi:2005gj},
the consequences of the positive Higgs signal were studied \cite{recent}.
It was shown \cite{Christensen:2012ei} that for the excess of $\gamma\gamma$ events as the result of a SM-like Higgs boson in the mass range
\begin{equation}
\sim 125 \gev \pm 2 \gev ,
\label{eq:mh1}
\end{equation}
the MSSM Higgs parameters split into two distinct regions. One region (the ``non-decoupling" region \cite{HHaber,Boos:2002ze}) has $m_A\lesssim130$~GeV.
In this region, the light CP-even Higgs $h^0$ and the CP-odd state $A^0$ are nearly mass degenerate and close to $\sim m_{Z}$, while the charged state $H^\pm$ and the heavy CP-even state $H^0$ are heavier and close to 125~GeV.  In the other region (the ``decoupling" region), the light CP-even Higgs $h^0$ has a mass around 125~GeV, while all the other Higgs bosons are heavy and decoupled \cite{HHaber,Gerard:2007kn}.
%
In particular, if certain channels (such as $W^+W^-$) indeed turn out to be smaller than the SM expectation, it would then
imply that the other Higgs bosons do not decouple and again suggests the non-decoupling region.
%
%
%
As a result, as pointed out in \cite{Christensen:2012ei}, if the other Higgs bosons are light and fall into the non-decoupling region, they may be more accessible at the LHC than previously thought.

The current experimental studies for Higgs bosons will continue to improve by analyzing the leading channels
\begin{equation}
gg \to h^{0}, H^{0}, A^{0}\quad\mbox{and}\quad pp\to t\bar t\ {\rm with}\ t\to H^{\pm} b,
\label{ggh}
\end{equation}
as well as the standard electroweak production processes
\begin{equation}
pp\to W^{\pm} h^{0} (H^{0}),\ Z h^{0} (H^{0}),\ \  {\rm and}\ \ q\bar q h ^{0} (H^{0}).
\label{VH}
\end{equation}
All of these processes have a substantial dependence on the parameters of the MSSM.
In this paper, we would like to emphasize the potential importance of the electroweak production of pairs of Higgs bosons and explore their observability. In particular, the processes
\begin{equation}
pp\to H^{\pm} A^{0},\ H^{+}H^{-},
\label{gauge}
\end{equation}
are via pure electroweak gauge interactions and are independent of the MSSM parameters except for their masses in contrast to the processes in Eqs.~(\ref{ggh}) and (\ref{VH}).
%
Additionally, there may be sizable contributions from the processes
\begin{equation}
pp\to H^{\pm} h^{0},\ A^{0} h^{0},
\label{mssm}
\end{equation}
in the low-mass non-decoupling region which, however, do depend on the MSSM parameters and, thus, may be used to distinguish them.

The rest of the paper is organized as follows. In Sec.~\ref{MSSMH}, we recall the Higgs sector of the MSSM and present the parameter choices for our study. In Sec.~\ref{HPair}, we calculate the signal cross section for the Higgs pair production channels $H^\pm A^0$, $H^\pm h^0$, $H^+H^-$ and $A^0h^0$ and explore the observability of the signal over the SM background. We end this section with estimates for the integrated luminosity necessary for discovery of $H^\pm A^0$, $H^\pm h^0$ and $H^+H^-$ at the 8 TeV and 14 TeV LHC.
We summarize our results in Sec.~\ref{Sum}.


\section{MSSM Higgs sector and Parameters}
\label{MSSMH}

In the MSSM, the two SU(2)$_{L}$ Higgs doublets result in five physical Higgs bosons after electroweak symmetry breaking: two CP-even states $h^{0}$ and $H^{0}$, one CP-odd state $A^{0}$ and a pair of charged scalars $H^{\pm}$.
At tree level, the masses of the Higgs bosons and the mixing angle of the CP-even states ($\alpha$)
can be expressed in terms of two parameters \cite{Gunion:1989we,Djouadi:2005gj},
conventionally chosen as
the mass of the CP-odd Higgs $(m_{A})$ and the ratio of the two vacuum expectation values ($\tan\beta=v_u/v_d$):

\begin{eqnarray}
&&
m_{h^0, H^0}^2 = \frac{1}{2} \left( (m_A^2 + m_Z^2) \mp \sqrt{(m_A^2 - m_Z^2)^2 + 4 m_A^2 m_Z^2 \sin^22 \beta} \right),\\
&&
m_{H^\pm}^2 = m_A^2 + m_W^2, \quad \cos^{2}(\beta-\alpha) = {
{\mh^{2} (m_{Z}^{2} - \mh^{2} )} \over {\ma^{2} (\mH^{2} - \mh^{2})}
 } .
\end{eqnarray}
We will call the CP-even Higgs boson that couples to $\ww$ and $ZZ$ more strongly the ``Standard Model-like'' (SM-like) Higgs.

The CP-even Higgs boson masses receive significant radiative corrections
due to the large top-quark Yukawa coupling and, potentially, from the large mixing of the left and right top squarks.  Inclusion of the leading one-loop terms from the top sector yields the
masses \cite{Djouadi:2005gj,Carena:1995wu}
%
\begin{eqnarray}
m_{h^0,H^0}^2 &=& \frac{1}{2}\left(m_A^2+m_Z^2+\epsilon\right)^2\left[1\mp\sqrt{1-4\frac{m_Z^2m_A^2\cos^22\beta+\epsilon\left(m_A^2\sin^2\beta+m_Z^2\cos^2\beta\right)}{\left(m_A^2+m_Z^2+\epsilon\right)^2}}\right],\\
\epsilon &=& \frac{3m_t^4}{2\pi^2v^2\sin^2\beta}\left[\ln\left(\frac{M_S^2}{m_t^2}\right)+\frac{\tilde{A}_t^2}{2M_S^2}\left(1-\frac{\tilde{A}_t^2}{6M_S^2}\right)\right] ,\\
\tilde{A}_t &=& A_t-\mu\cot\beta ,
\end{eqnarray}
where $M_S=(m_{\tilde{t}_1}+m_{\tilde{t}_2})/2$ is the arithmetic average of the stop masses, $A_t$ is the stop trilinear coupling and $\mu$ is the Higgs mixing parameter in the Superpotential.
For the mass calculations in these formulas, there are uncertainties of order a few GeV coming from higher loop orders, as well as from the uncertainties in $m_t$, $\alpha_s$, etc..

In Ref.~\cite{Christensen:2012ei}, we studied the 6-dimensional parameter space in the ranges
\begin{eqnarray}
\nonumber
& 3 < \tan\beta < 55, \quad
 50\ {\rm GeV}< m_A < 500\ {\rm GeV}, \quad 100\gev <  \mu < 1000\ {\rm GeV}, &\\
&   100\gev < M_{3SU}, M_{3SQ} < 2000\ {\rm GeV}, \quad
    {-4000\ \gev}< A_t < 4000\  {\rm GeV}. &
\label{eq:para}
\end{eqnarray}
%
We will scan in the same region, however, since we are interested in the non-decoupling region, we will focus our scan by reducing the range of $m_A$ and $A_t$ to
\begin{equation}
 95\ {\rm GeV}< m_A < 130\ {\rm GeV}  , \quad 0<A_t<4000\ \mbox{GeV}.
 \label{eq:ma}
\end{equation}
We performed our scan by using the FeynHiggs 2.8.6
package \cite{Degrassi:2002fi}
to calculate the mass spectrum, couplings and other SUSY parameters.
We calculated the Higgs pair cross sections in CalcHEP \cite{Belyaev:2012qa} using the couplings given by FeynHiggs.
We used
HiggsBound 3.6.1beta \cite{Bechtle:2008jh}
to check the exclusion constraints from LEP2 \cite{LEP2H}, the Tevatron \cite{CDFD0} and the LHC \cite{ATLASH}.
We further checked the updated exclusions from the LHC \cite{LHCJuly4} which were not included in the HiggsBound package.
We generated a large random data sample that passed these constraints for this study.

%


\section{Search for non-SM-like MSSM Higgs at the LHC}
\label{HPair}

\begin{figure}[tb]
\begin{center}
\includegraphics[scale=1,width=8.1cm]{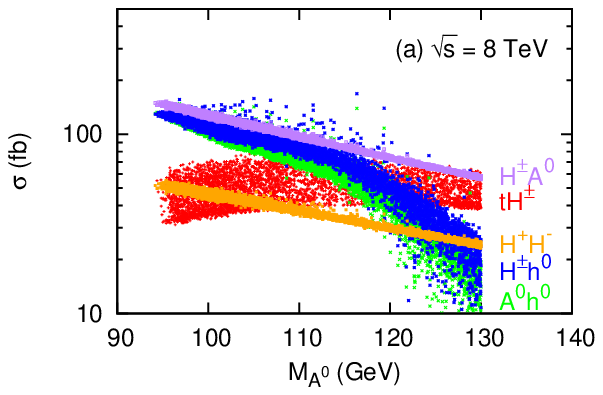}
\includegraphics[scale=1,width=8.1cm]{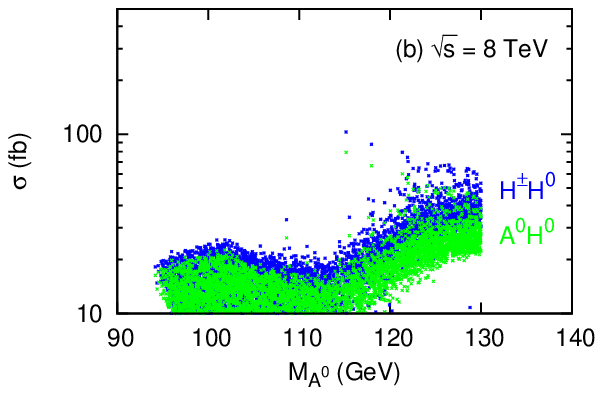}
\includegraphics[scale=1,width=8.1cm]{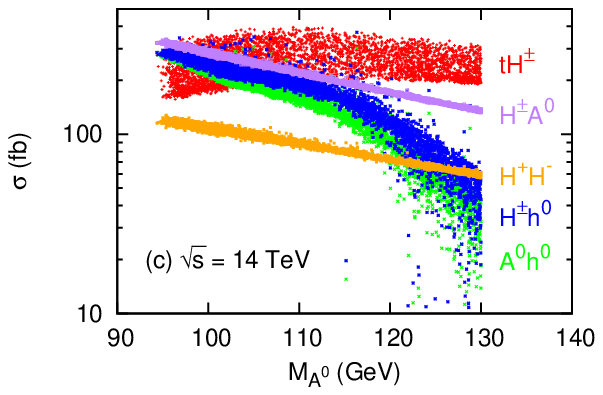}
\includegraphics[scale=1,width=8.1cm]{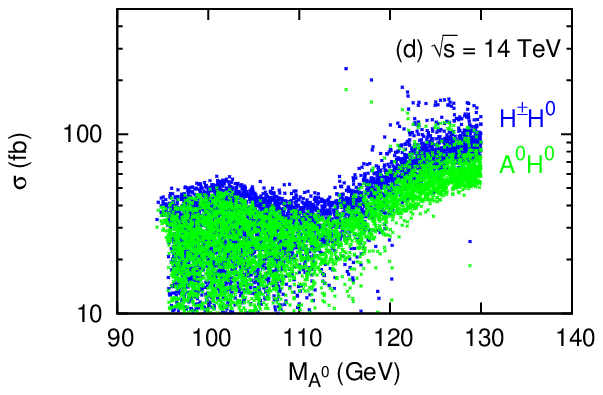}
\end{center}
\caption{Production cross sections for MSSM Higgs boson pairs versus $\ma$ scanned over  Eqs.~(\ref{eq:para}) and (\ref{eq:ma}) for the non-SM-like (a and c) and the SM-like (b and d) Higgs at the LHC with 8 TeV (a and b) and 14 TeV (c and d) C.M. energy.}
\label{totcs8-14}
\end{figure}

While the searches for Higgs bosons at the LHC will continue to improve by probing the standard processes in Eqs.~(\ref{ggh}) and (\ref{VH}), we point out the importance of Higgs pair production in the non-decoupling region.
%
We are interested in non-SM-like MSSM Higgs pair production through the Drell-Yan processes
\begin{eqnarray}
&& q\bar{q}'\to W^{\pm \ast} \to H^\pm \A , H^\pm\h , \label{eq:pair 1}\\
&& q\bar{q}\to Z^\ast/ \gamma^\ast \to H^+ H^- , \ \  q\bar{q}\to Z^\ast \to \A \h \label{eq:pair 2}.
\end{eqnarray}
We show scatter plots for the production cross sections versus $\ma$ scanned over the parameters in Eqs.~(\ref{eq:para}) and (\ref{eq:ma}) in Figs.~\ref{totcs8-14}(a) and (b)
 for 8 TeV C.M.~energy and in Figs.~\ref{totcs8-14}(c) and (d) for  14 TeV C.M.~energy. We see that the total cross section for the leading pair production channel $H^\pm \A$ is about 60$-$180 fb in the mass range of current interest at 8 TeV and approximately doubles at 14 TeV. The channel $H^+H^-$ is  roughly a factor of three smaller.
These two channels are independent of MSSM parameters except for the EW gauge coupling and their physical masses, as evidenced from the narrow bands in the scatter plots. The production rates for the other two channels $H^{\pm}\h$ and $\A\h$ are similar to that of $H^{\pm}\A$ at low $\ma$, and then drop below $H^{+}H^{-}$ near $\ma \sim 125$ GeV.
%
For comparison, we have also shown the QCD-EW production $gb\to t H^{\pm}$. It is interesting to note that the leading EW Higgs pair production channels at 8 TeV for the low mass $\ma \sim 95$ GeV are significantly larger than the $t H^{\pm}$ production, and they become comparable at 14 TeV.
In Figs.~\ref{totcs8-14}(b) and (d), we show the cross sections for the sub-leading processes $H^\pm \H$  and $\A \H$
for the C.M.~energies of 8 and 14 TeV, respectively. These two processes are strongly dependant on the Higgs sector parameters but are complementary to those of   $H^\pm \h$  and $\A \h$.

%
%

\begin{figure}[!tb]
\begin{center}
\includegraphics[scale=1,width=8.1cm]{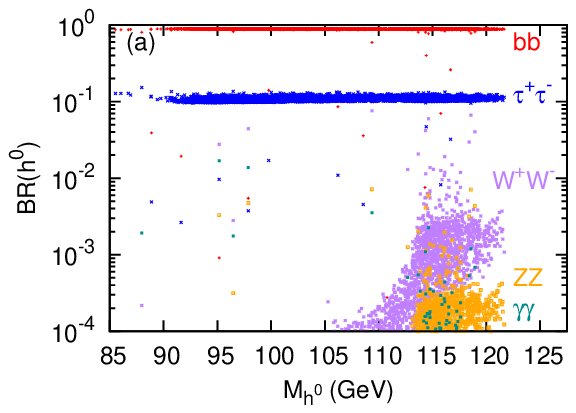}
\includegraphics[scale=1,width=8.1cm]{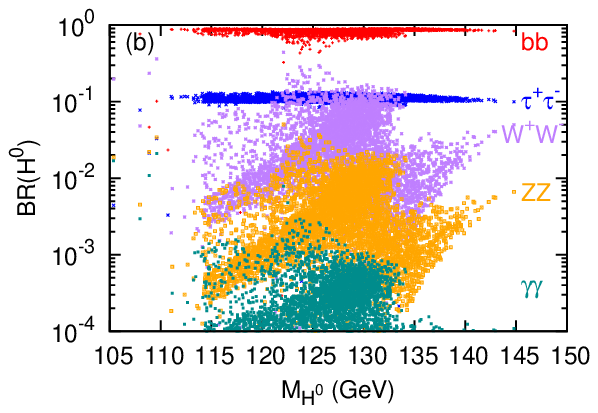}\\
\includegraphics[scale=1,width=8.1cm]{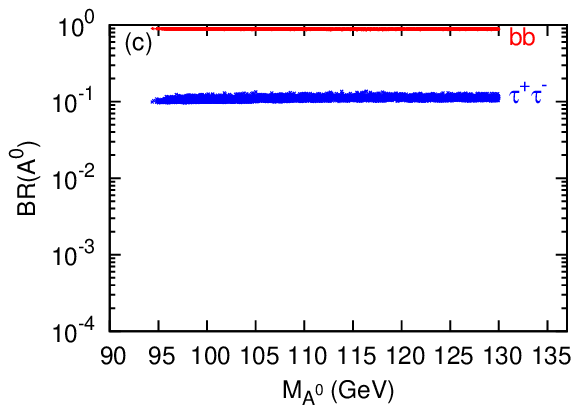}
\includegraphics[scale=1,width=8.1cm]{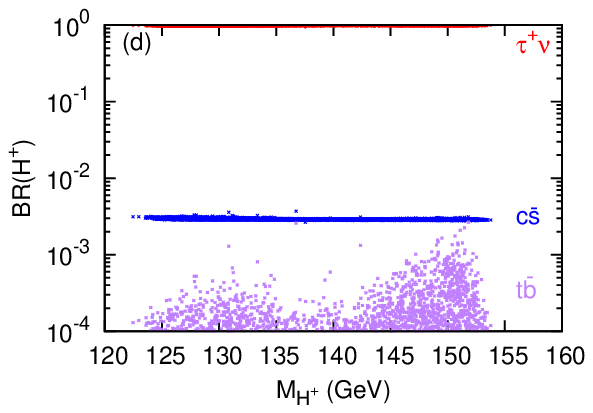}
\end{center}
\caption{Branching fractions for the MSSM Higgs bosons (a) $\h$, (b) $\H$, (c) $\A$, and (d) $H^{\pm}$ scanned over Eqs.~(\ref{eq:para}) and (\ref{eq:ma}).}
\label{BRs}
\end{figure}
We next show the decay branching fractions in the parameter ranges of Eqs. (\ref{eq:para}) and (\ref{eq:ma}) in Fig.~\ref{BRs}.
We see that the dominant decays $\A, \h, \H \to b\bar b$ are near $90\%$, $H^{\pm} \to \tau^{\pm} \nu$ is near $100\%$, and the sub-dominant decays
$\A, \h, \H \to \tautau$ are near $10\%$. The SM-like Higgs boson $\H$ has further accessible channels scattered over a large range (due to their dependence on the MSSM parameters), such as $\ww$ with about $10\%$, $ZZ$ for a few percent, and $\gamma\gamma$ at the level of $10^{-3}$.

In the following, we analyze the non-SM-like MSSM Higgs pair production channels in Eqs. (\ref{eq:pair 1}) and (\ref{eq:pair 2}) at the 8 TeV and 14 TeV LHC. For the sake of illustration, we analyze the signals as well as the corresponding backgrounds based on the benchmark point given in Table~\ref{bmp}. We will generalize the study when we evaluate the sensitivity for the signal observation of the Higgs pair production in Sec.~\ref{sec:Sensitivity}. We apply an overall next-to-leading (NLO) QCD $K$-factor of 1.3 to all Higgs pair production channels via $q\bar q$ annihilation \cite{QCD}.
We have not taken into account the kinematical dependence of the K-factor for different distributions, either for the signal nor for the backgrounds. We consider the crude estimate justifiable at least for the signal since we are far away from the kinematical regions with high invariant mass or high rapidity boost, where the QCD effects become significant.

\begin{table}[tb]
\begin{tabular}{|c|c|c|c|c|c|c|c|c|c|c|}
\hline
$M_{SUSY}$ & $M_{3SQ}$ & $M_{3SU}$ & $A_t$ & $\mu$ & $m_A$ & $\tan\beta$ & $\cos(\beta-\alpha)$ & $m_{h^0}$ & $m_{H^0}$ & $m_{H^\pm}$\\ \hline
3 TeV & 1.86 TeV & 1.63 TeV & 2.0 TeV & 0.49 TeV & 96 GeV & 13 & $-0.95$ & 94 GeV & 125 GeV & 126 GeV \\ \hline
\end{tabular}
\caption{MSSM benchmark point in the non-decoupling region satisfying the bounds from LEP, Tevatron and LHC.}
\label{bmp}
\end{table}


We focus on the $\tau$ and $b$ final states. We adopt the $\tau$ hadronic decays to take  advantage of spin correlation for the final state hadrons \cite{KH,Boos:2005ca}.
%
The branching fractions for the $\tau$ decays are $BR(\tau^\pm\to \pi^\pm \nu_\tau)=0.11$ and $BR(\tau^\pm\to \rho^\pm \nu_\tau)=0.25$.
The $b$-jet tagging efficiency at the LHC is taken to be $\epsilon_{b}=70\%$ \cite{ATLASH}.
%
We employ the following basic acceptance cuts for the event selection
\begin{eqnarray}
p_T(h_\tau,b)\geq 20 \ {\rm GeV}; \ |\eta(h_\tau,b)|<2.4; \ \Delta R_{h_\tau b}, \Delta R_{h_\tau h_\tau}, \Delta R_{bb} \geq 0.4,
\label{ah+-basic}
\end{eqnarray}
where $h_\tau$ denotes the charged pion or rho. We will also impose a cut on missing transverse energy $\cancel{E}_T$, which will be optimized according to a specific final state process.
%
To simulate the detector effects, we smear the hadronic energy by a Gaussian distribution whose width is parameterized as~\cite{smearing}
\begin{eqnarray}
{\Delta E\over E}={a_{had}\over \sqrt{E/{\rm GeV}}} \oplus b_{had}, \ \ a_{had}=100\%,
b_{had}=5\%.
\end{eqnarray}

We use Madgraph5 and Madevent to generate signal and background events~\cite{MGME}, and Tauola interfaced with Pythia to simulate tau lepton decay carrying polarization information~\cite{Tauola}.

\subsection{$H^\pm \A \to  \tau^\pm \nu_\tau\ b\bar{b}$}

As discussed in the last section, the $H^{\pm} \A$ channel is one of the leading signal modes in the non-decoupling region. The signal consists of one tau lepton and missing energy from $H^\pm$ decay, plus two $b$ jets from $A^0$ decay. The leading SM backgrounds to this channel are
\begin{eqnarray}
b\bar{b}W^\pm \to b\bar{b} \tau^\pm \nu,\ {\rm and}  \ W^{*} \to \bar{b}t\ (b\bar{t})\to \bar{b}bW^\pm\to \bar{b}b \tau^\pm \nu ,
\end{eqnarray}
where the contributions from $g,\ \gamma,\ Z \to b\bar b$ are included in the first process. The second process is the $s$-channel single top production.
Other top-quark production also yields a large background
\begin{eqnarray}
&& qg \to q\bar{b}t(b\bar{t})\to j\bar{b}bW^\pm\to j\bar{b}b \tau^\pm \nu, \\
&& t\bar{t}\to b\bar{b}W^+W^- \to b\bar{b}\tau^\pm \ell^\mp \nu \bar{\nu} \ (\ell=e,\mu),
\end{eqnarray}
where the first one is the single top production from $Wg$ fusion, and the second is the QCD $t\bar t$ production. These processes have additional jet or lepton activity and can thus be reduced by vetoing extra jets and leptons with
\begin{eqnarray}
{\rm veto:}\ \ p_T(j)>30 \ {\rm GeV},\ |\eta(j)|<4.9\  {\rm or}\ p_T(\ell)>7 \ {\rm GeV}, \ |\eta(\ell)|<3.5.
\label{veto}
\end{eqnarray}
The QCD corrections to the background processes have also been included and the
next-to-leading (NLO) $K$-factors of order 2 (2.7), 0.9 (0.9) and 1.5 (1.63) for $b\bar{b}W^\pm$~\cite{NLObbw}, $bt$~\cite{NLObt} and $t\bar{t}$~\cite{NLOtt} at 8 (14) TeV LHC are adopted.

\begin{figure}[tb]
\begin{center}
\includegraphics[scale=1,width=7.5cm]{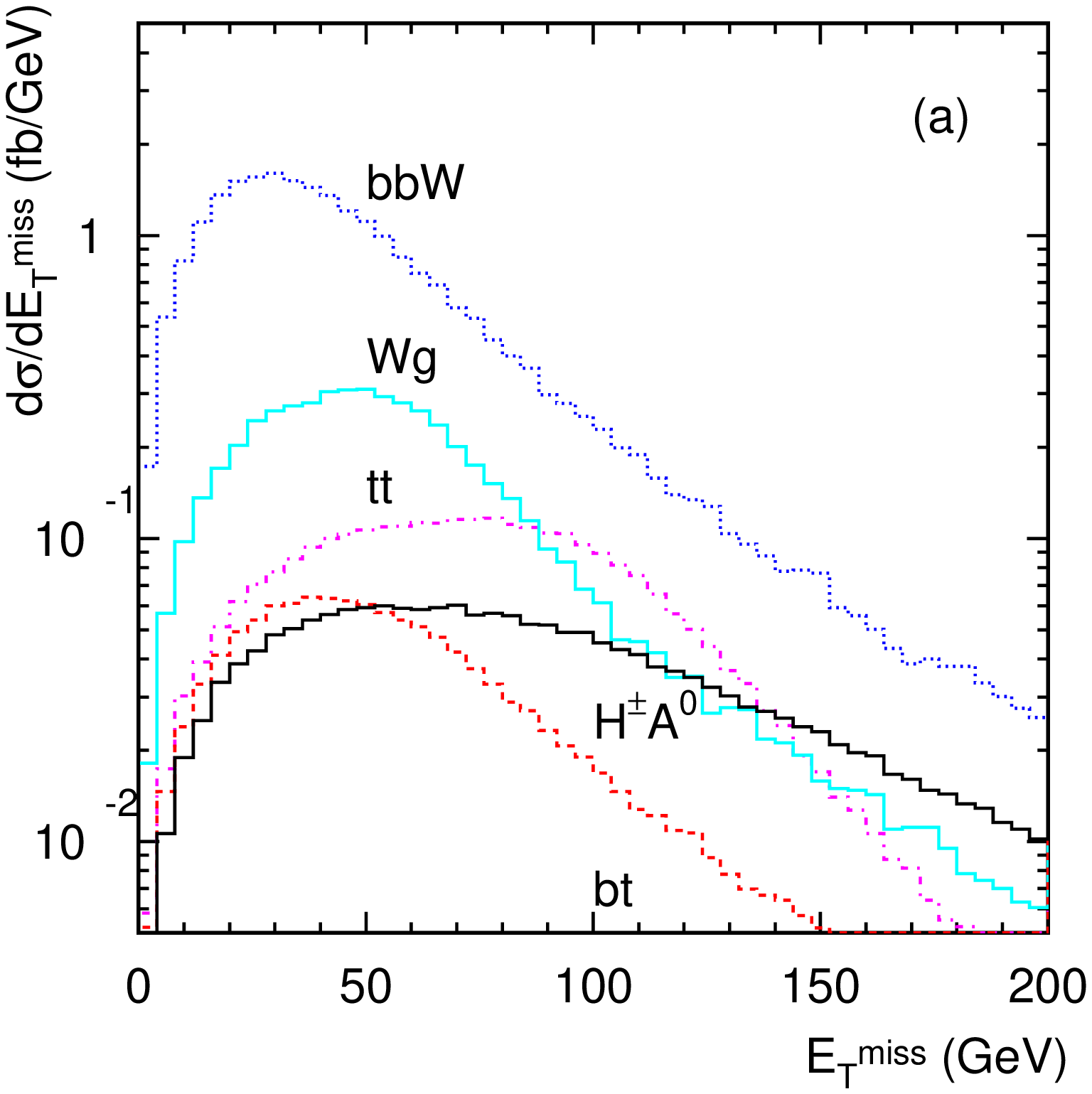}
\includegraphics[scale=1,width=7.5cm]{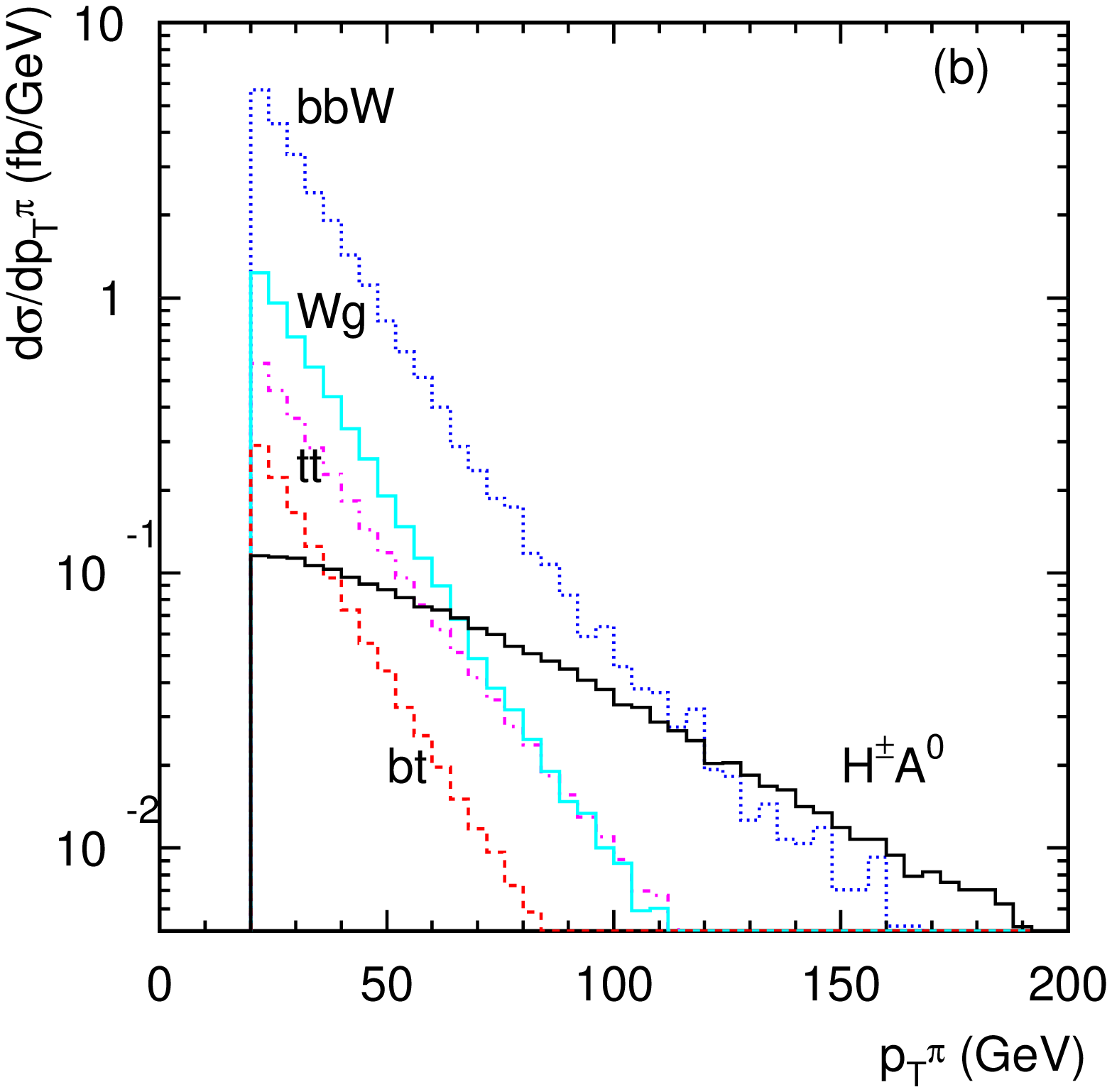}\\
\includegraphics[scale=1,width=7.5cm]{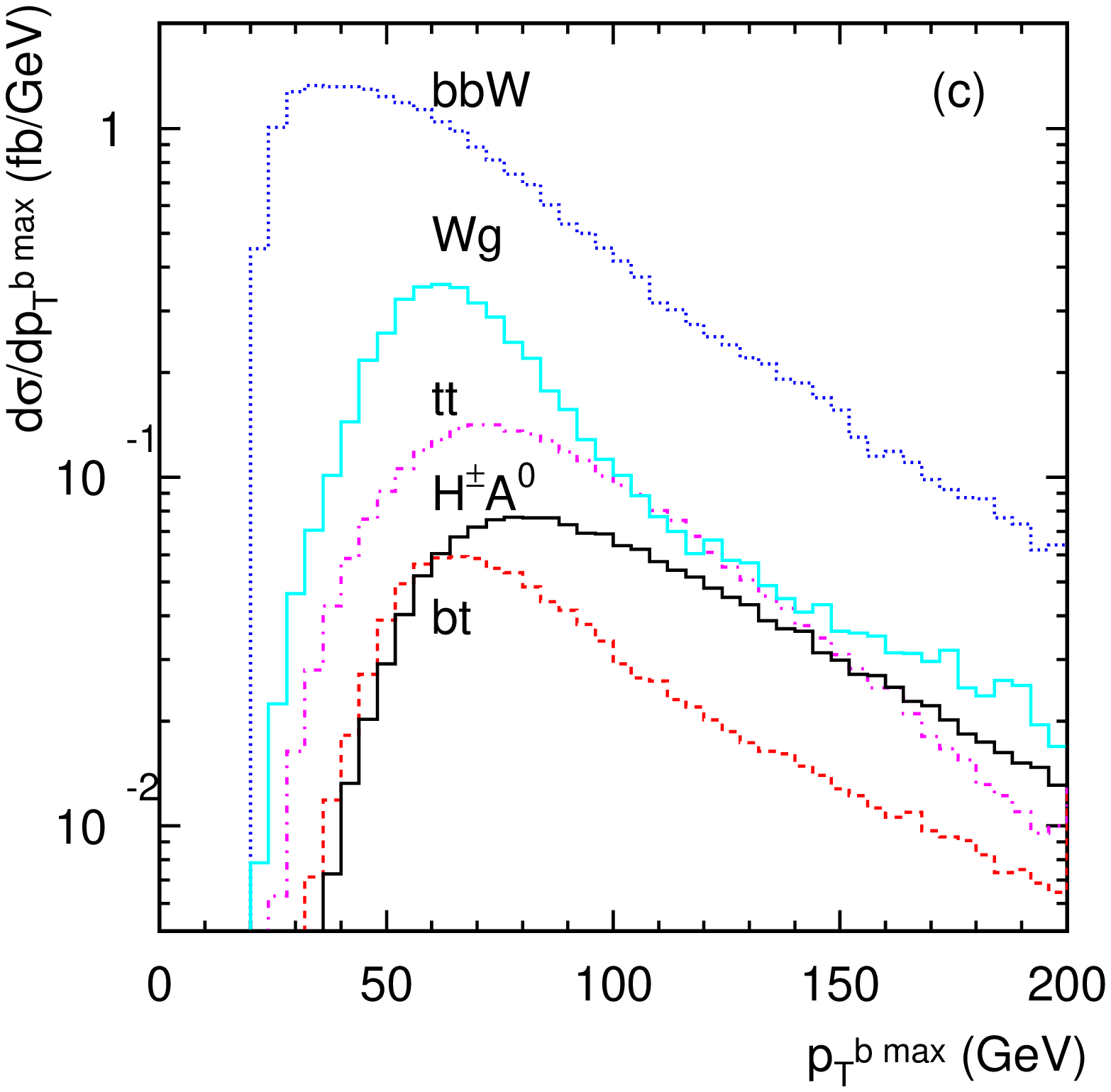}
\includegraphics[scale=1,width=7.5cm]{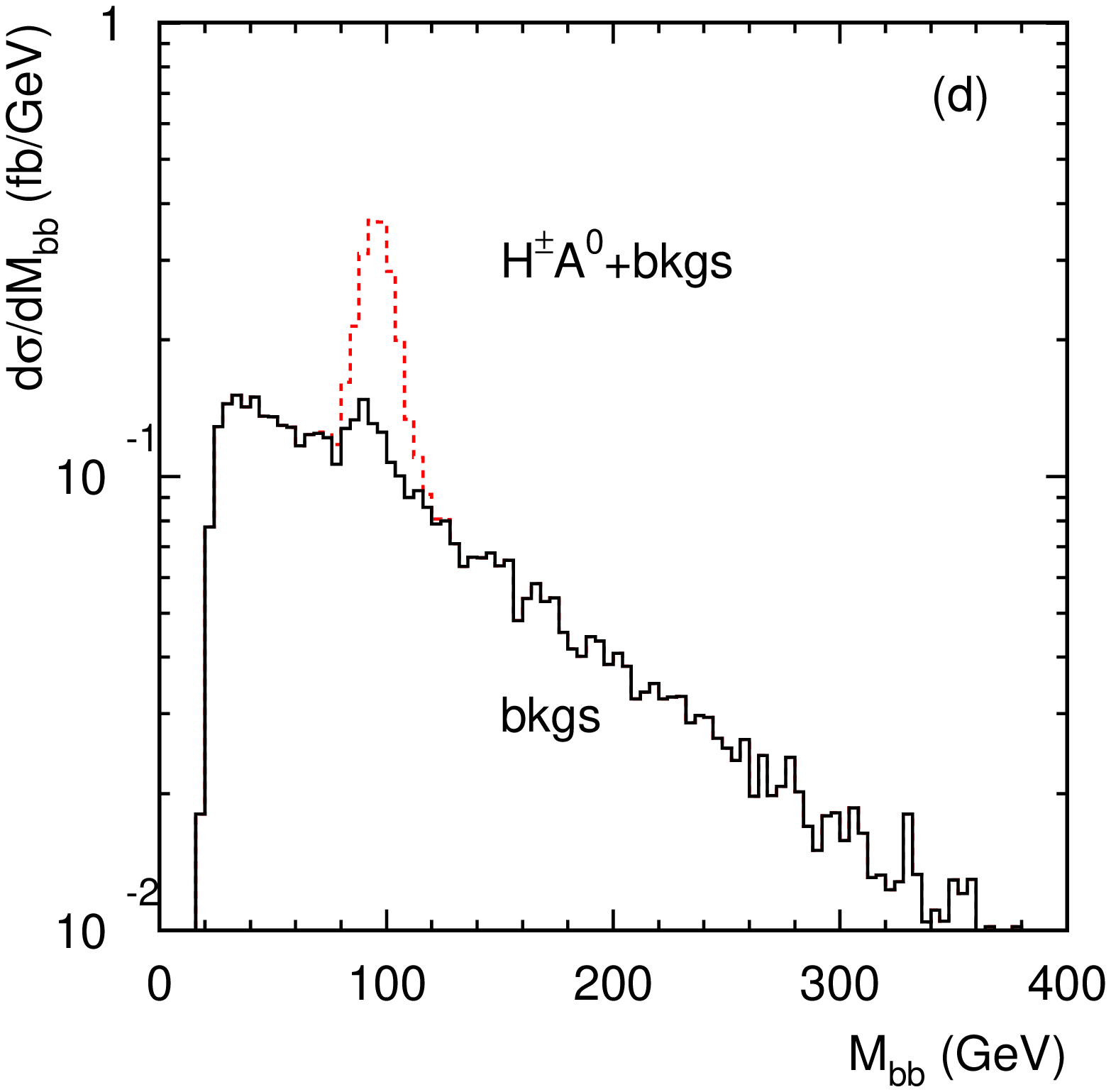}
\end{center}
\caption{The differential cross section distributions of the signal $H^\pm\A$ and backgrounds versus (a) $\cancel{E}_T$, (b) $p_{T\pi}$, (c) $p_{Tb}^{\rm max}$, and (d) $M_{b\bar{b}}$ at the 14 TeV LHC.}
\label{ah+-sigbkg14}
\end{figure}

Although the background rates are very large to begin with, the signal and background kinematics are quite different. We first study the decay mode $\tau^\pm \to \pi^\pm \nu_\tau$ and take into account
the tau decay into $\rho^\pm$ and $\nu_\tau$ later on. We display the distributions of signal and backgrounds at the 14 TeV LHC after the basic cuts shown
in Eq.~(\ref{ah+-basic}) in Fig.~\ref{ah+-sigbkg14}, for (a) missing transverse energy $\cancel{E}_T$, (b) transverse pion momentum $p_{T\pi}$ and (c) transverse momentum for the harder b-tagged jet $p_{Tb}^{\rm max}$.
We first note that the signal has a harder $\cancel{E}_T$ spectrum than the background.
This is from a smeared-out distribution around the Jacobean peak at $p_{T\nu} \sim \mHpm /2.$
Furthermore, the signal also has a harder $p_{T\pi}$ spectrum compared to the background.
This is a well-known result of spin correlation in the $\tau$ decay.   For the $H^{+}$ signal, the left-handed $\tau^{+}$ decays to a right-handed $\bar\nu_{\tau}$, causing the $\pi^{+}$ to preferentially move along the $\tau^{+}$ momentum direction~\cite{KH,CP}. In contrast, the $\tau^{+}$ coming from a $W^+$ decay is right-handed  which has the opposite effect on the $\pi^+$.
We thus tighten the selection cuts by imposing
\begin{eqnarray}
\cancel{E}_T>40 \ {\rm GeV}, \ \ p_T(\pi)>40 \ {\rm GeV}.
\end{eqnarray}
This helps reduce the background significantly.
The invariant mass of the two $b$-jets $M_{b\bar{b}}$ after all cuts mentioned above
is shown in Fig.~\ref{ah+-sigbkg14}(d). The $Z\to b\bar b$ contribution is visible near $M_{Z}$. When estimating the signal observability near the $A^0$ resonance, we take a mass window for the invariant mass of $b\bar b$ of
\begin{equation}
80 \ {\rm GeV}<M_{b\bar{b}}<110\ {\rm GeV}.
\label{eq:mass}
\end{equation}



\begin{table}[tb]
\begin{tabular}{|c|c|c|c|c|c|}
\hline
events with 8 (14) TeV  & basic cuts+$\cancel{E}_T>40$ GeV & $p_T(\pi/\rho)>40$ GeV & $80 \ {\rm GeV}<M_{b\bar{b}}<110$ GeV
\\ \hline
$H^\pm A^0$ ($\pi$) & 47 (96) & 31 (66) & 29 (61)
\\
$H^\pm A^0$ ($\rho$)& 110 (225) & 70 (150) & 65 (140)
\\ \hline
$H^\pm h^0$ ($\pi$) & 44 (90) & 28 (63) & 26 (60)
\\
$H^\pm h^0$ ($\rho$) & 105 (210) & 65 (140) & 62 (135)
\\ \hline \hline
$b\bar{b}W^\pm$ ($\pi$) & 290 (760) & 75 (210) & 14 (37)
\\
$b\bar{b}W^\pm$ ($\rho$) & 1150 (2900) & 340 (920) & 66 (165)
\\ \hline
$bt$ ($\pi$) & 25 (49) & 6.1 (12) & 0.8 (1.5)
\\
$bt$ ($\rho$) & 100 (190) & 29 (60) & 4.2 (7.5)
\\ \hline
$Wg$ ($\pi$) & 77 (220) & 18 (55) & 2.6 (8.3) \\
$Wg$ ($\rho$) & 300 (850) & 88 (270) & 15 (43) \\
\hline
$t\bar{t}$ ($\pi$) & 30 (140) & 9.6 (48) & 1.6 (7.9)
\\
$t\bar{t}$ ($\rho$) & 117 (550) & 47 (230) & 7.9 (38)
\\ \hline  \hline
$S/B$ ($H^\pm A^0$, $\pi$) & 0.11 (0.08) & 0.29 (0.2) & 1.5 (1.1)
\\
$S/B$ ($H^\pm A^0$, $\rho$) & 0.066 (0.05) & 0.14 (0.1) & 0.7 (0.55)
\\ \hline
$\sqrt{LL}$ ($H^\pm A^0$, $\pi$) & 2.2 (2.8) & 2.8 (3.5) & 5.6 (7.2)
\\
$\sqrt{LL}$ ($H^\pm A^0$, $\rho$) & 2.7 (3.3) & 3.0 (3.8) & 6.1 (8.1)
\\ \hline
$S/B$ ($H^\pm h^0$, $\pi$) & 0.1 (0.077) & 0.26 (0.19) & 1.4 (1.1)
\\
$S/B$ ($H^\pm h^0$, $\rho$) & 0.063 (0.047) & 0.13 (0.095) & 0.67 (0.53)
\\ \hline
$\sqrt{LL}$ ($H^\pm h^0$, $\pi$) & 2.1 (2.6) & 2.6 (3.4) & 5.1 (7.1)
\\
$\sqrt{LL}$ ($H^\pm h^0$, $\rho$) & 2.5 (3.1) & 2.8 (3.6) & 5.9 (7.9)
\\ \hline
\end{tabular}
\caption{The number of signal ($H^\pm A^0$ and $H^\pm h^0$) and background events expected with $\tau^\pm\to \pi^\pm \nu_\tau$ or $\rho^\pm \nu_\tau$ after kinematic cuts at the 8 (14) TeV LHC with a luminosity of 15 (15) fb$^{-1}$.}
\label{ah+-events}
\end{table}

The coupling of $W^\mp h^0H^\pm$ is proportional to $\cos(\beta-\alpha)$ which, as illustrated in Table \ref{bmp}, is $\sim-1$ in the non-decoupling region. For the present illustrative parameters (see Table \ref{bmp}), the production rate of $H^\pm\h$ is comparable to that of $H^\pm\A$ and their signals are exactly the same in this scenario. We include this contribution and apply the same kinematic cuts described above on the $H^\pm \h \to \tau^\pm \nu_\tau\ b\bar{b}$ production.
We summarize the signals $H^\pm\A$ and $H^\pm\h$ together with the background events for $\tau^\pm\to \pi^\pm \nu_\tau$ and $\rho^\pm \nu_\tau$ after kinematic cuts in consecutive steps at 8 (14) TeV LHC with an integrated luminosity of 15 (15) fb$^{-1}$ in Table~\ref{ah+-events}.  For our calculation of the significance here and below, we used the log likelihood method
\begin{equation}
LL(B,S)=2\left[(B+S)\ln\left(\frac{B+S}{B}\right)-S\right],
\end{equation}
where $B$ is the background expectation and $S$ is the signal expectation.
We see that we could achieve a signal-to-background ratio of the order of unity after our cuts.
By combining the $\pi$ and $\rho$ channels at 8 TeV, we find that it is possible to reach $5\sigma$ sensitivity for the $\Hpm \A$ signal with 6.0 fb$^{-1}$ and for the $\Hpm \h$ signal with 6.7 fb$^{-1}$. At the 14 TeV LHC, one could reach a $5\sigma$ sensitivity for each individual hadronic channel (either $\pi$ or $\rho$) with as little as 7.5 fb$^{-1}$.

\subsection{$H^{+}H^- \to \tau^+ \nu_\tau \ \tau^- \bar \nu_\tau$}

\begin{figure}[tb]
\begin{center}
\includegraphics[scale=1,width=7.5cm]{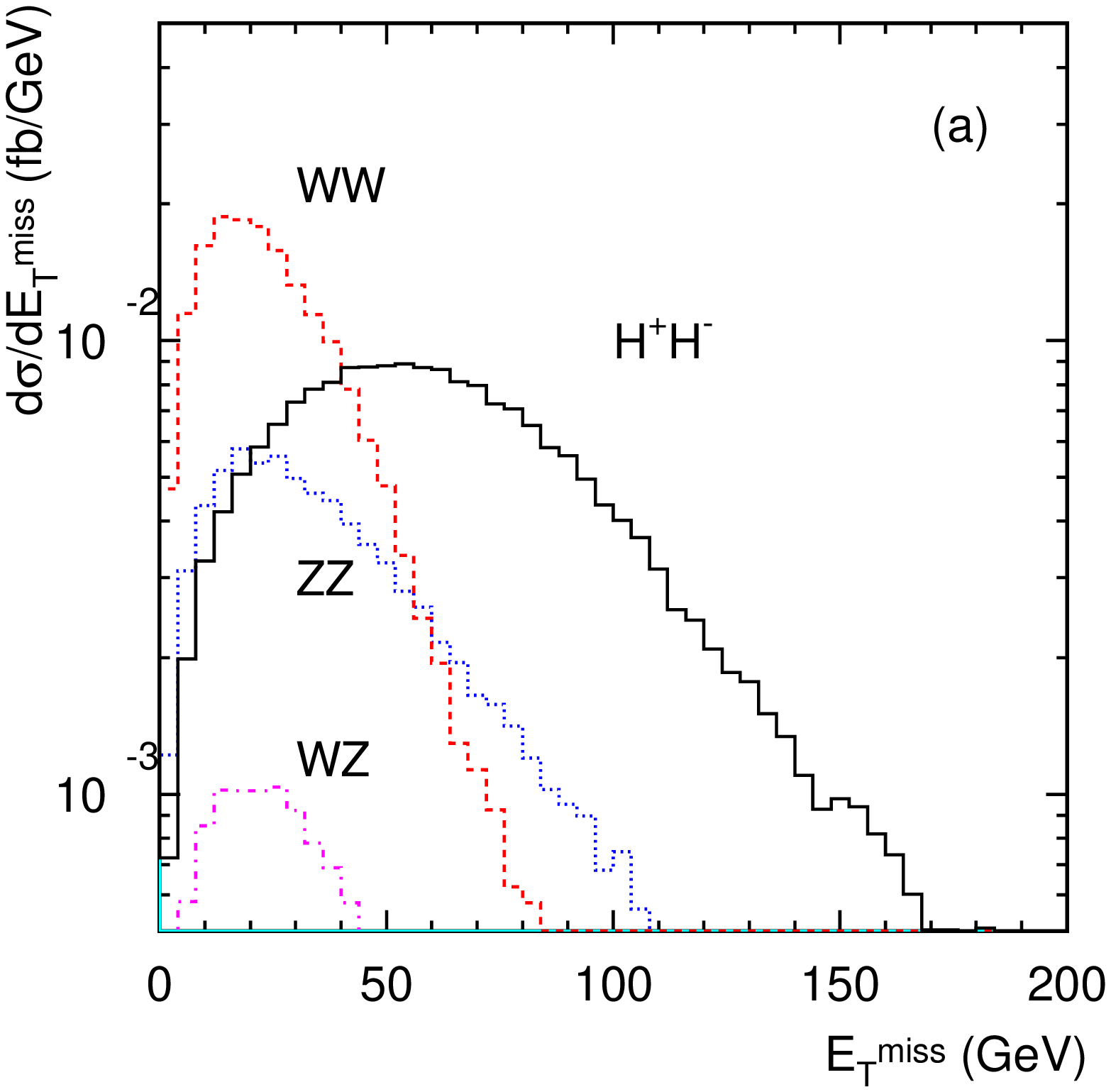}
\includegraphics[scale=1,width=7.5cm]{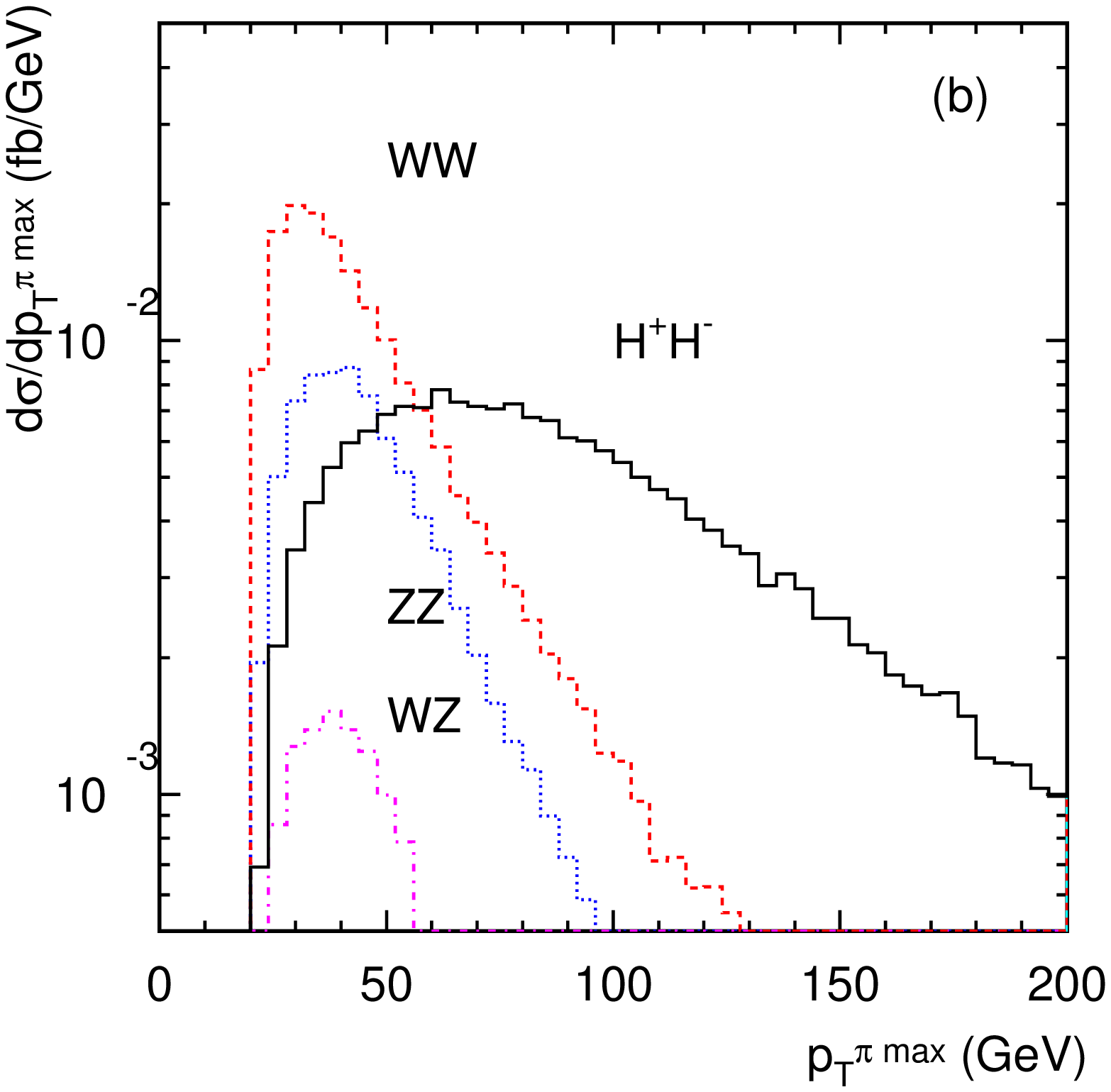}
\end{center}
\caption{The differential cross section distributions of the signal $H^+H^-$ and backgrounds versus (a) $\cancel{E}_T$ and (b) $p_{T\pi}^{\rm max}$ at the 14 TeV LHC.}
\label{h+h-sigbkg14}
\end{figure}

The other model-independent Higgs channel from pure gauge interactions is $H^+H^-$ pair production. The leading decay channel is $H^+H^-\to \tau^+ \nu_\tau\ \tau^- \bar{\nu}_\tau$ with nearly a $100\%$ branching fraction.
The leading SM backgrounds are
\begin{eqnarray}
W^+W^-\to \tau^+\nu_\tau\ \tau^-  \bar{\nu}_\tau,\  \ ZZ\to \tau^+\tau^-\ \nu\bar{\nu},\ \
W^\pm Z\to \ell^\pm \nu_\ell \tau^+\tau^-, 
\end{eqnarray}
where the charged lepton from $W^\pm Z$ production is vetoed using the same requirement as in Eq.~(\ref{veto}).
We apply the $K$-factors of 1.5, 1.3 and 1.7 to the channels $WW$, $ZZ$ and $WZ$,
respectively~\cite{Dixon:1999di}. We first study the decay mode $\tau^\pm \to \pi^\pm \nu_\tau$.  The signal would thus be two opposite-sign charged pions plus missing energy. We employ the same basic cuts as given in Eq.~(\ref{ah+-basic}) and display the kinematical distributions of missing transverse energy $\cancel{E}_T$ and the transverse momentum of the hardest pion $p_{T\pi}^{\rm max}$
in Figs.~\ref{h+h-sigbkg14}(a) and (b), respectively. One can see that the spin correlation effects mentioned earlier tend to be more dramatic in this channel (in comparison with the $WW$ background) because the visible objects (two pions here) are purely from the polarized tau decays.
Therefore, we strengthen the cuts further as
\begin{eqnarray}
\cancel{E}_T > 50 \ {\rm GeV},\  \ p_{T\pi}^{\rm max} > 50 \ {\rm GeV}.
\end{eqnarray}
The number of events expected for the signal and backgrounds and the statistical significance at 8 (14) TeV with $\tau^\pm\to \pi^\pm \nu_\tau$ or $\rho^\pm \nu_\tau$ are shown in Table~\ref{h+h-events} after the cuts in consecutive steps.
We see that a signal-to-background ratio of about 1$-$3 is achievable. Combining the $\pi$ and $\rho$ channels at 8 TeV, one could reach a $5\sigma$ sensitivity for the $H^{+}H^{-}$ signal with about 20 fb$^{-1}$. At the 14 TeV LHC, one could reach a $5\sigma$ sensitivity with as little as 8.4 fb$^{-1}$.


\begin{table}[tb]
\begin{tabular}{|c|c|c|}
\hline
events with 8 (14) TeV & basic cuts+$\cancel{E}_T>50$ GeV & $p_T(\pi/\rho)>50$ GeV
\\ \hline
$H^+H^-$ ($\pi$) & 3.5 (7.6) & 3.2 (7.0)
\\
$H^+H^-$ ($\rho$) & 18 (39) & 16 (36)
\\ \hline \hline
$WW$ ($\pi$) & 0.52 (1.1) & 0.44 (0.97)
\\
$WW$ ($\rho$) & 12 (23) & 8.4 (17)
\\ \hline
$ZZ$ ($\pi$) & 0.77 (1.9) & 0.58 (1.3)
\\
$ZZ$ ($\rho$) & 5.7 (11) & 4.2 (9.0)
\\ \hline
$WZ$ ($\pi$) & 0.057 (0.16) & 0.043 (0.12)
\\
$WZ$ ($\rho$) & 0.37 (1.1) & 0.26 (0.80)
\\ \hline \hline
$S/B$ ($\pi$) & 2.6 (2.4) & 3.0 (2.9)
\\
$S/B$ ($\rho$) & 1.0 (1.1) & 1.2 (1.3)
\\ \hline
$\sqrt{LL}$ ($\pi$) & 2.3 (3.3) & 2.3 (3.4)
\\
$\sqrt{LL}$ ($\rho$) & 3.7 (5.7)  & 3.8 (5.9)
\\ \hline
\end{tabular}
\caption{The number of signal ($H^+H^-$) and background events expected with $\tau^\pm\to \pi^\pm \nu_\tau$ or $\rho^\pm \nu_\tau$ after kinematic cuts at the 8 (14) TeV LHC with a luminosity of 15 (15) fb$^{-1}$.}
\label{h+h-events}
\end{table}

\subsection{$A^0h^0 \to b\bar{b}\ \tau^+ \tau^{-} $}

\begin{figure}[tb]
\begin{center}
\includegraphics[scale=1,width=7.5cm]{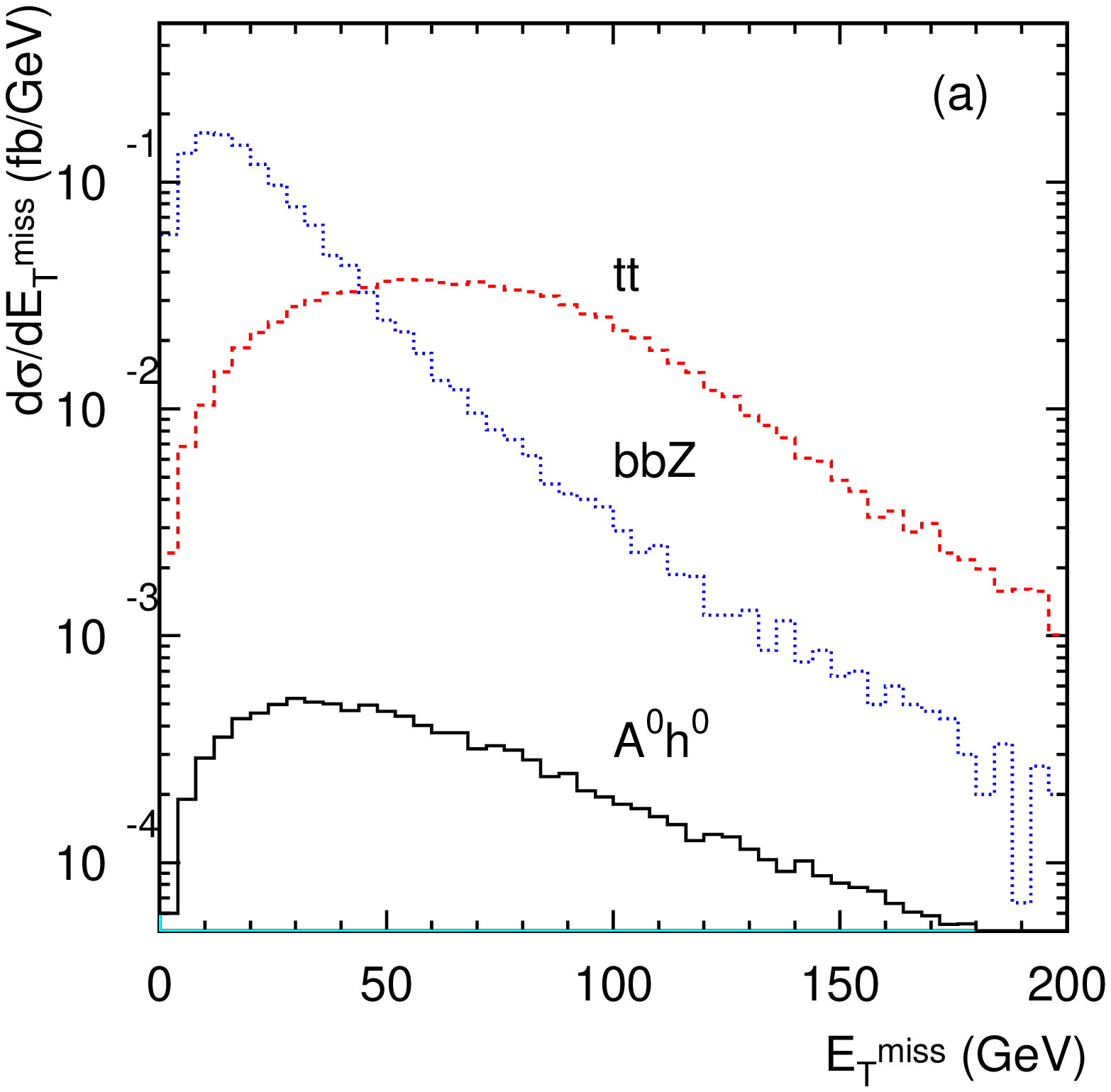}
\includegraphics[scale=1,width=7.5cm]{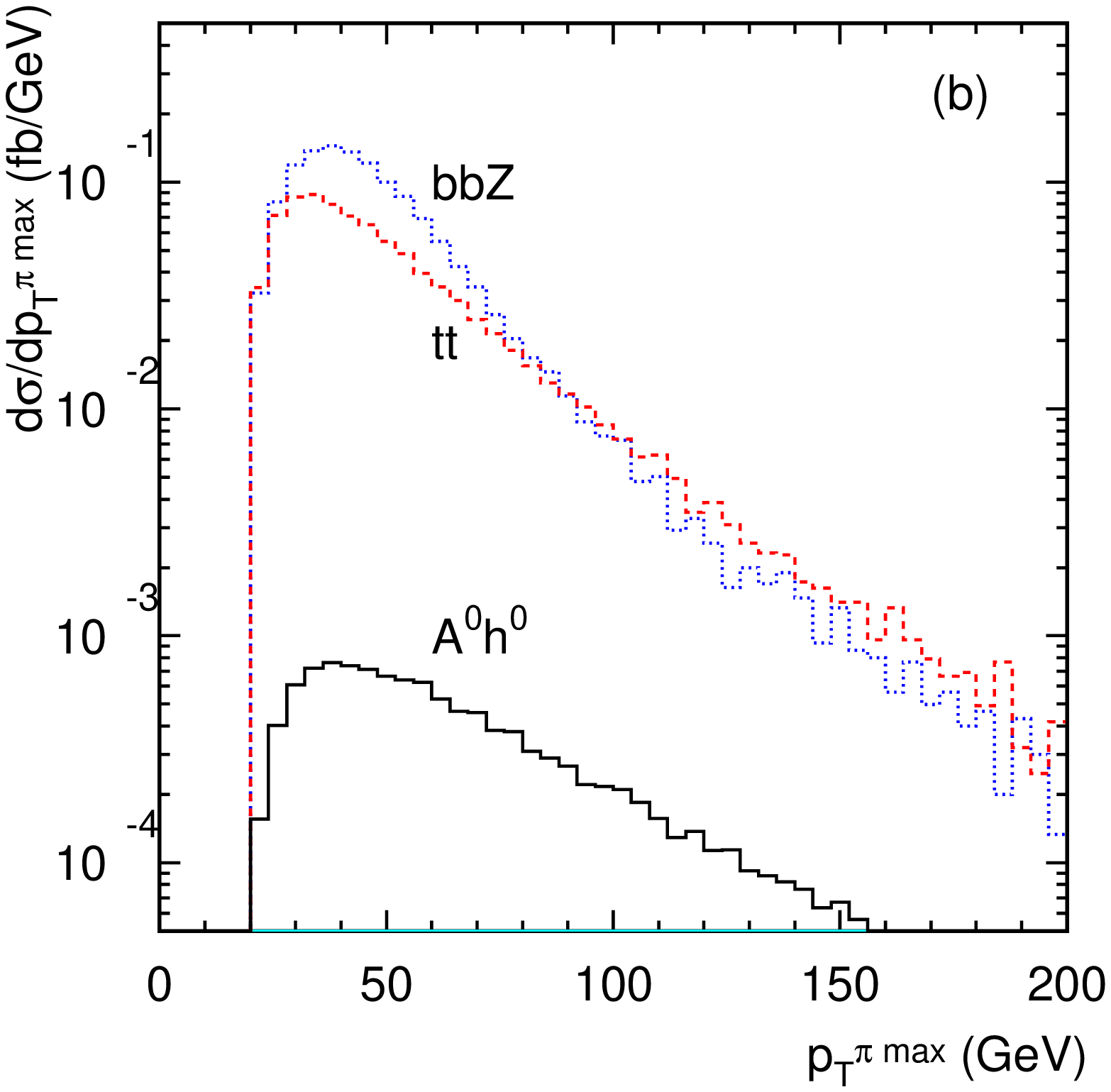}\\
\includegraphics[scale=1,width=7.5cm]{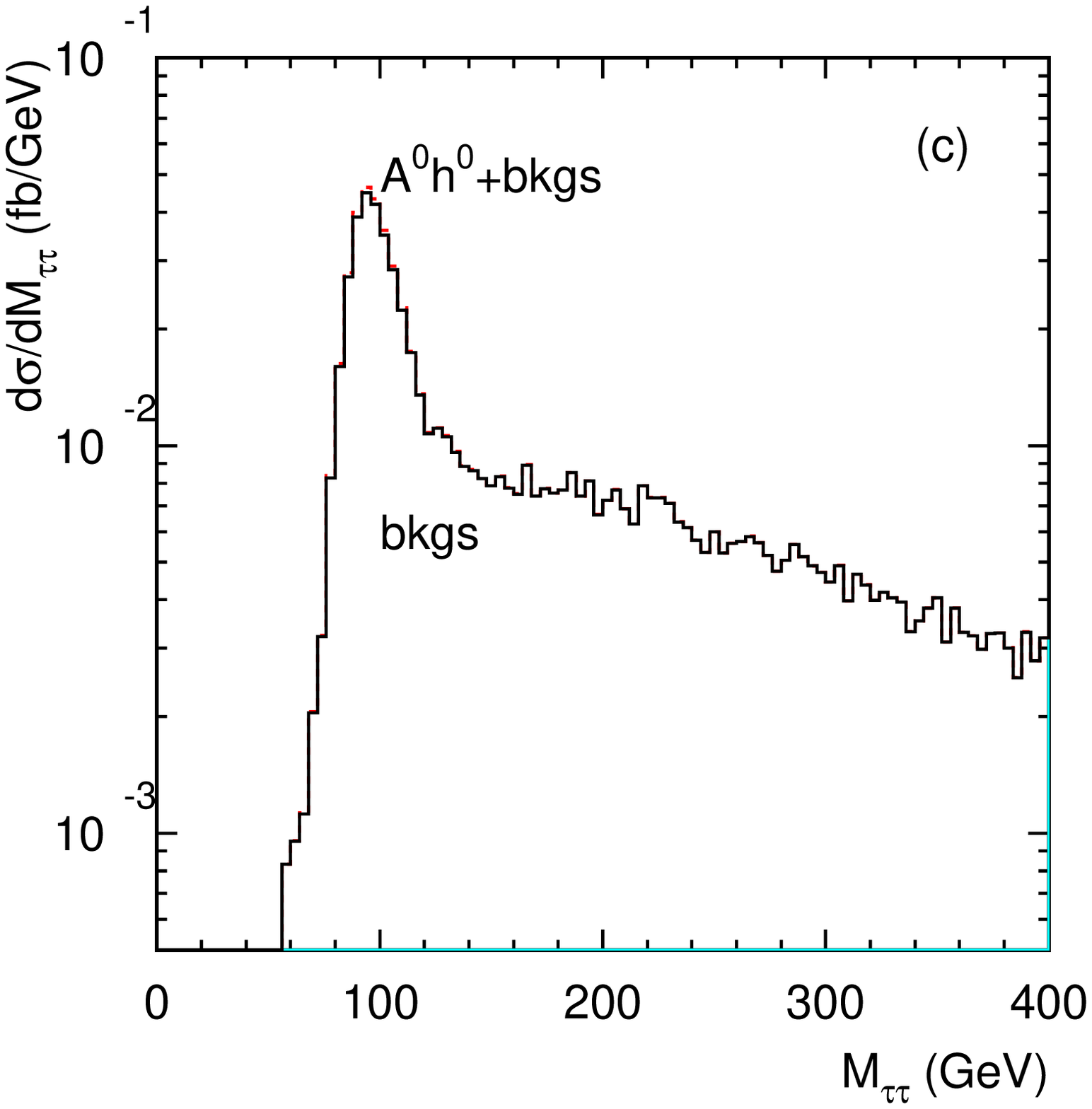}
\includegraphics[scale=1,width=7.5cm]{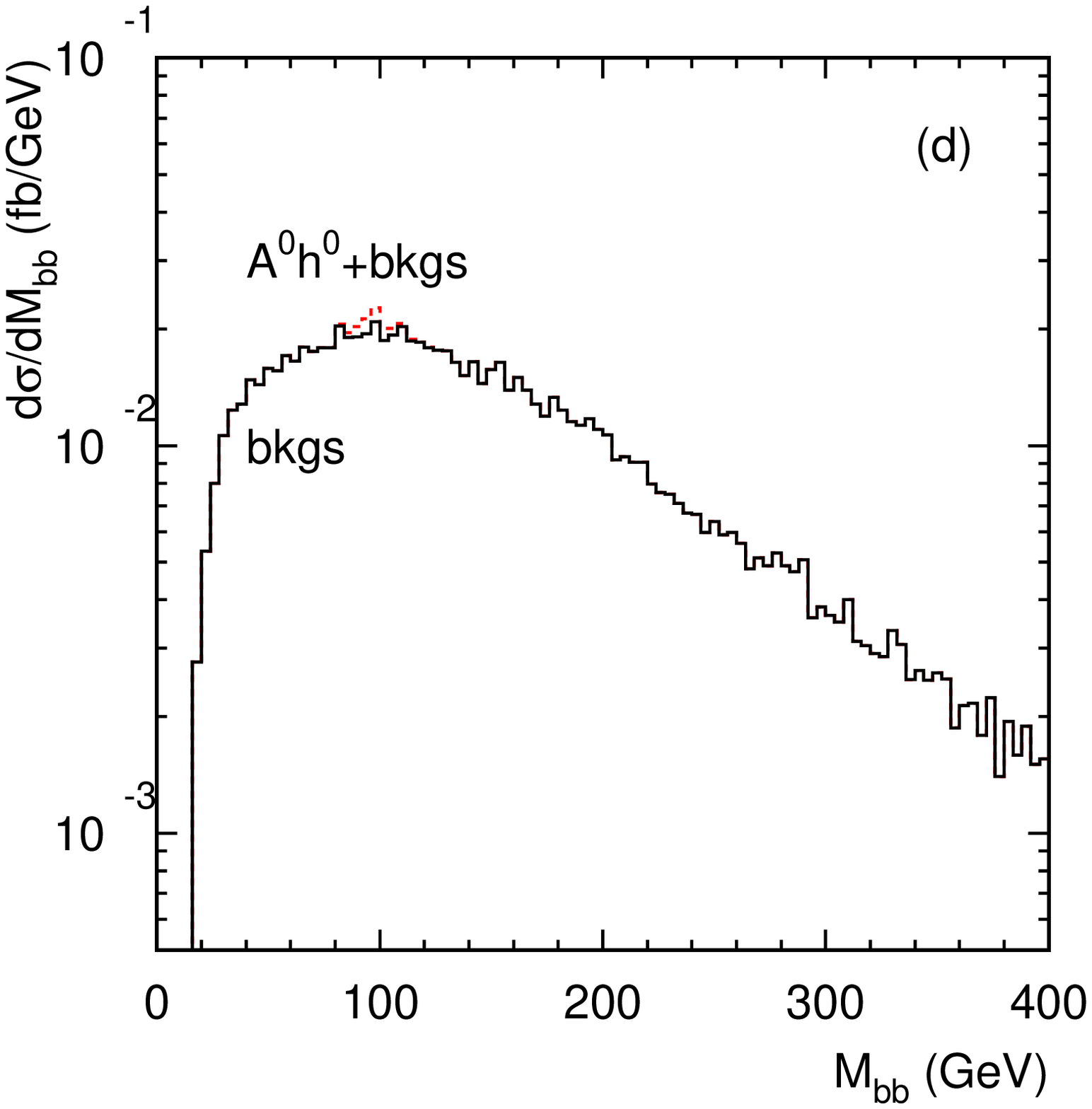}
\end{center}
\caption{The differential cross section distributions of the signal $h^0A^0$ and backgrounds versus (a) $\cancel{E}_T$, (b) $p_{T\pi}^{\rm max}$, (c) $M_{\tau\tau}$, and (d) $M_{b\bar{b}}$ at the 14 TeV LHC.}
\label{h1h3sigbkg14}
\end{figure}

As seen in Figs.~\ref{totcs8-14}(a) and (b), the other potentially important channel for the Higgs pair production in the low mass non-decoupling region is $q\bar{q}\to A^0\h$.
The coupling of $Zh^0A^0$ is  proportional to $\cos(\beta-\alpha)\sim -1$ and sizeable.
The leading signal, after decay, is $A^0\h \to b\bar{b}b\bar{b}$ which, however, would be overwhelmed by a huge QCD background. Thus, we consider the cleaner but subleading signal, namely two $b$-jets plus two opposite sign tau leptons,  $A^0 \h \to b\bar{b}\tau^+\tau^-$  with a $BR(h^0(A^0)\to \tau^+\tau^-)\approx 10\%$.
The $\tau$'s produced in these signal events are quite energetic, with an energy of approximately half the Higgs boson mass. Each missing neutrino will be approximately collinear with the direction of a corresponding charged pion. In this approximation, we take the missing neutrinos' momentum as
\begin{eqnarray}
\overrightarrow{p}({\rm missing})=\kappa_1\overrightarrow{p}(\pi_1)+\kappa_2\overrightarrow{p}(\pi_2),
\end{eqnarray}
where the proportionality constants $\kappa_1$ and $\kappa_2$ can be determined from the missing energy measurement as long as the two charged tracks are linearly-independent.

The dominant SM backgrounds to this channel are
\begin{eqnarray}
b\bar{b}Z\to b\bar{b} \tau^+\tau^-,\quad \ t\bar{t}\to b\bar{b}W^+W^-\to b\bar{b}\tau^+\tau^-\nu\bar{\nu}.
\end{eqnarray}
The NLO QCD $K$-factors for  $b\bar{b}Z$ are again included as 1.7 (2.2) for the 8 (14) TeV LHC~\cite{Frederix:2011qg}.
%
The distributions of missing transverse energy $\cancel{E}_T$ and transverse momentum of the hardest pion $p_{T\pi}^{\rm max}$ after applying the same basic cuts as in Eq.~(\ref{ah+-basic})
are shown in Figs.~\ref{h1h3sigbkg14}(a) and (b), respectively for the 14 TeV LHC and decay mode $\tau^\pm \to \pi^\pm \nu_\tau$. Due to the complex nature of the kinematics, these distributions do not present dramatic differences between the signal and backgrounds. We thus modestly strengthen the cuts as
\begin{eqnarray}
\cancel{E}_T > 30 \ {\rm GeV},\  \ p_{T\pi}^{\rm max} > 30 \ {\rm GeV}.
\end{eqnarray}
The reconstructed invariant mass distributions of the two tau's $M_{\tau\tau}$ and the two b's $M_{b\bar{b}}$ after all cuts described above are shown in Figs.~\ref{h1h3sigbkg14}(c) and (d)  for the 14 TeV LHC.
In estimating the signal statistical sensitivity, we take a mass windows for both as in Eq.~(\ref{eq:mass}) and
\begin{equation}
80 \ {\rm GeV} < M_{\tau\tau} < 110 \ {\rm GeV}.
\label{eq:M_tt cut}
\end{equation}
We could only reach a signal-to-background ratio of approximately 1:7. The signal rate is also low. With an integrated luminosity of 100 fb$^{-1}$ at 14 TeV, one would have only a handful events for the pion mode, and about 20 events for the rho mode. We thus conclude that the neutral Higgs pair production of $\A\h$ would not be a feasible channel for the MSSM Higgs pair search.

\subsection{Sensitivity\label{sec:Sensitivity}}

\begin{figure}[tb]
\begin{center}
\includegraphics[scale=1,width=8cm]{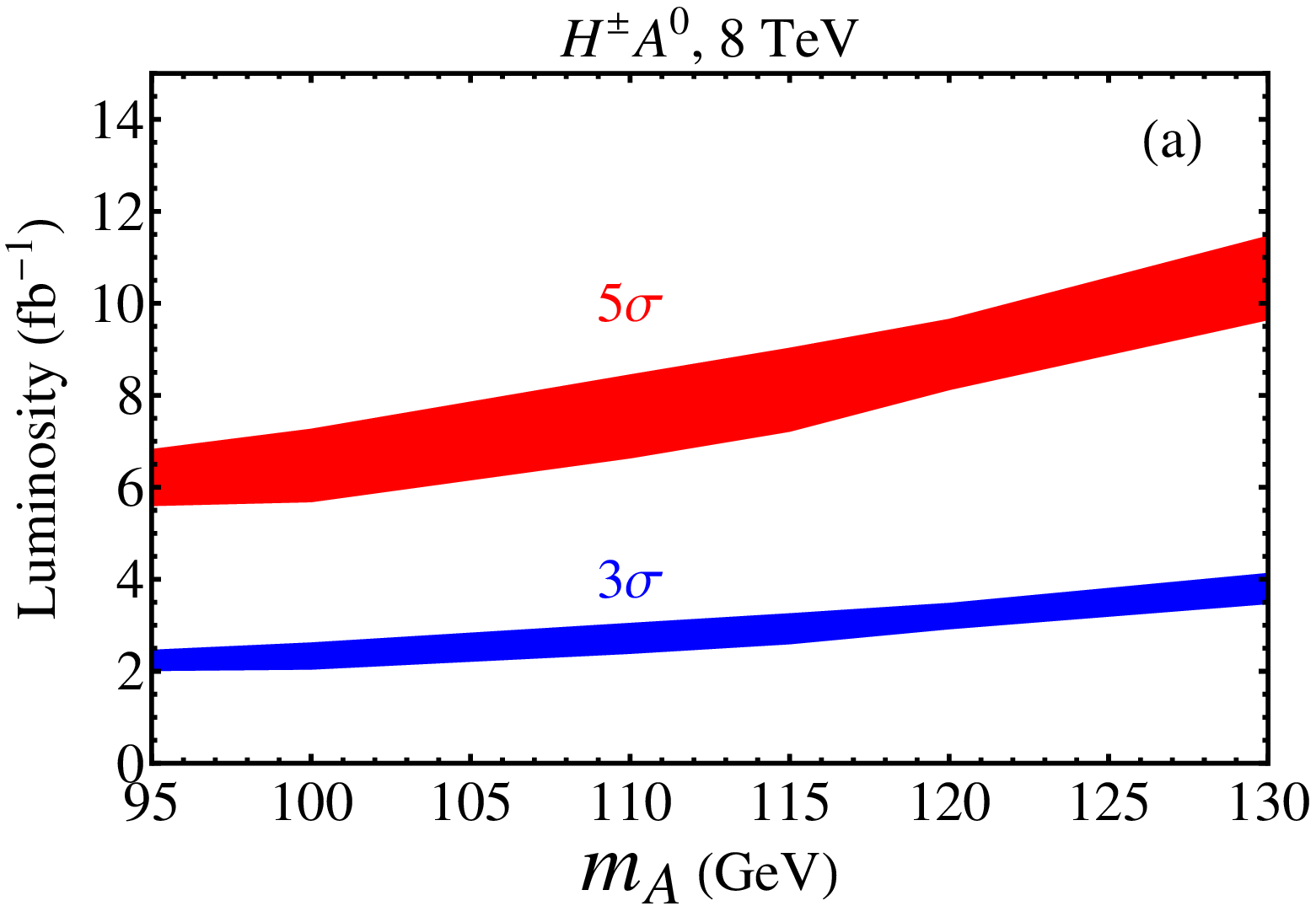}
\includegraphics[scale=1,width=8cm]{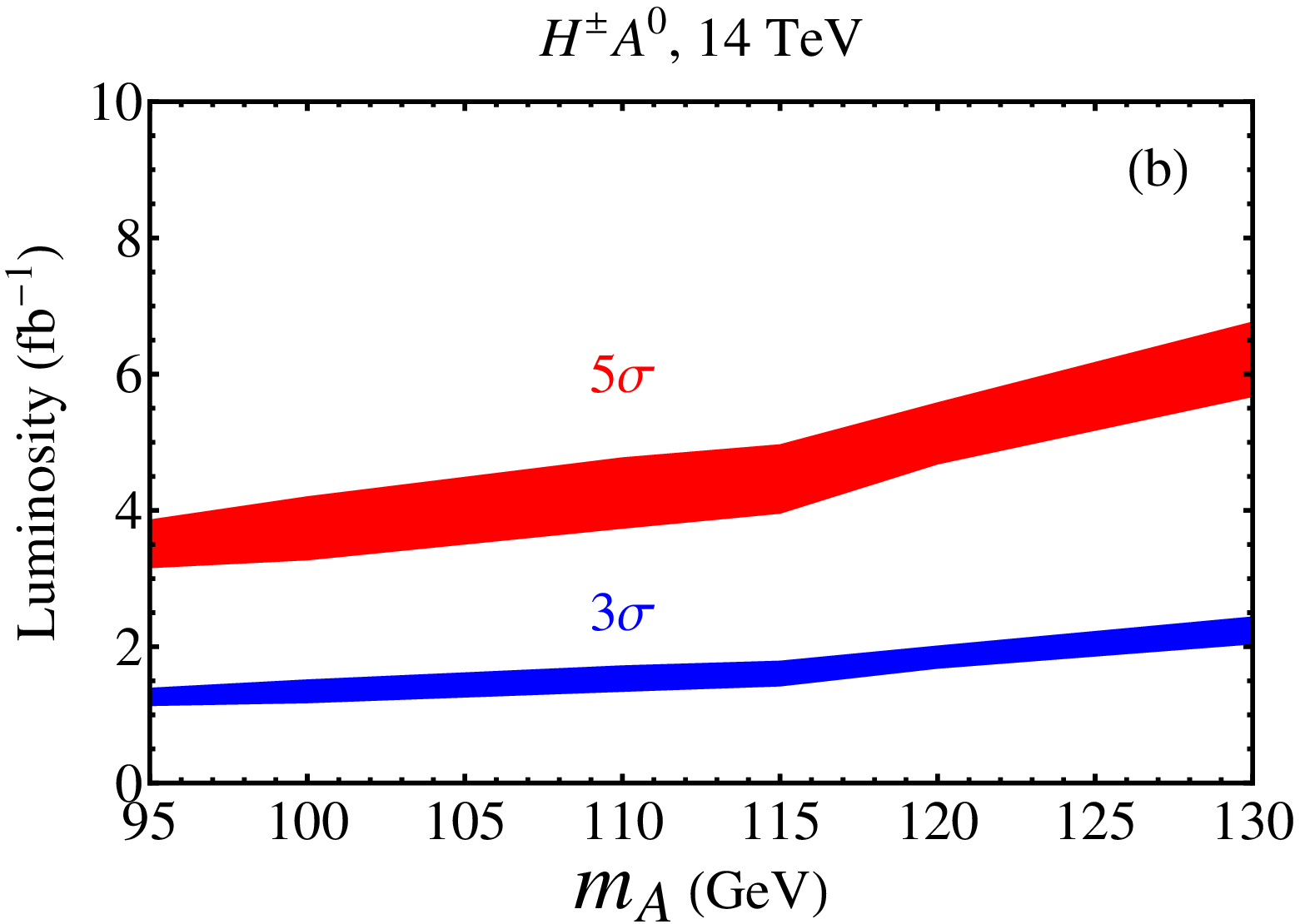}
\end{center}
\caption{Luminosity needed for $3\sigma$ (band on the bottom) and $5\sigma$ (band on the top) sensitivity as a function of $m_{A}$ for $H^\pm A^0$ at (a) 8 and (b) 14 TeV LHC.}
\label{ah+-sen}
\end{figure}

\begin{figure}[tb]
\begin{center}
\includegraphics[scale=1,width=8cm]{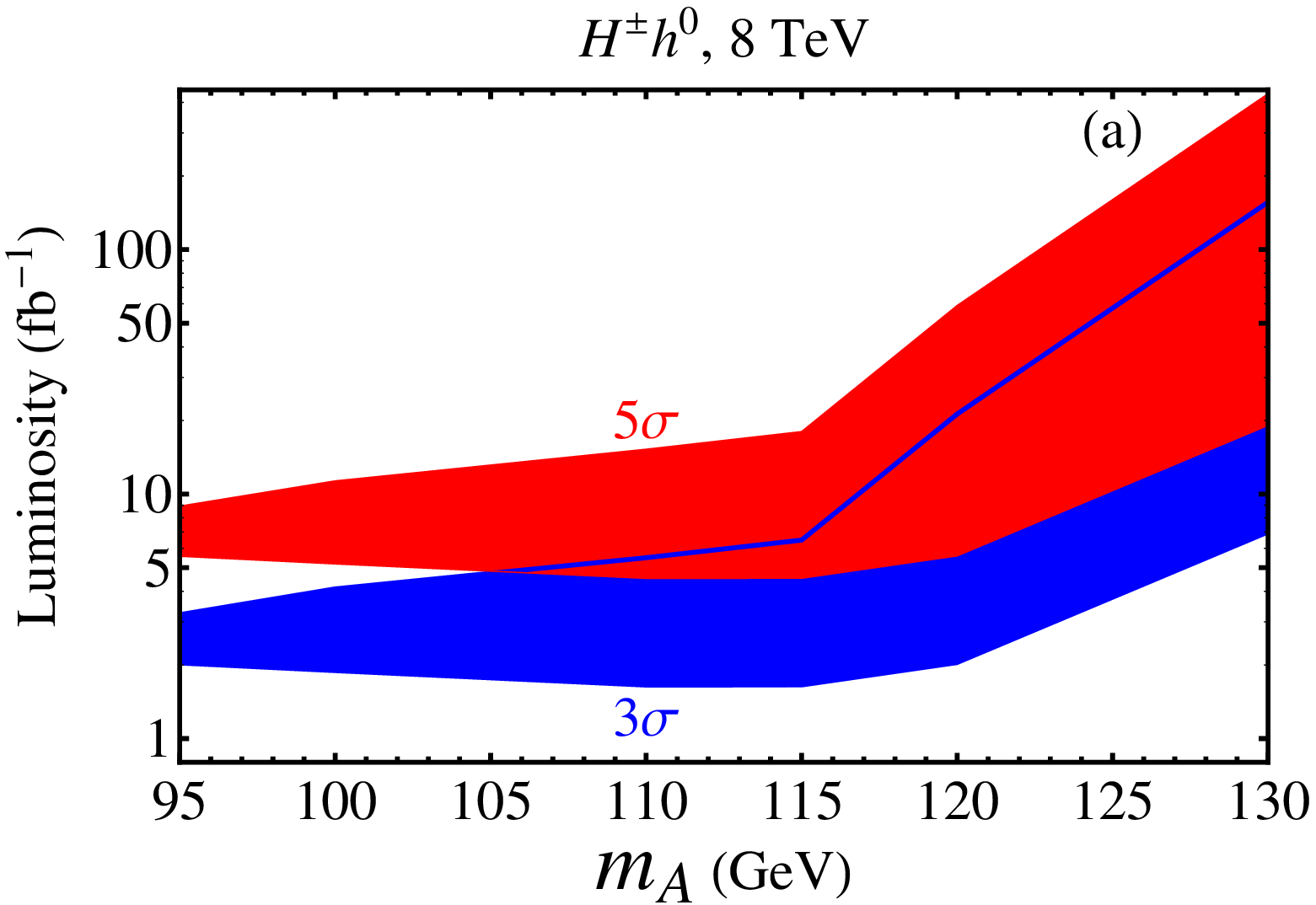}
\includegraphics[scale=1,width=8cm]{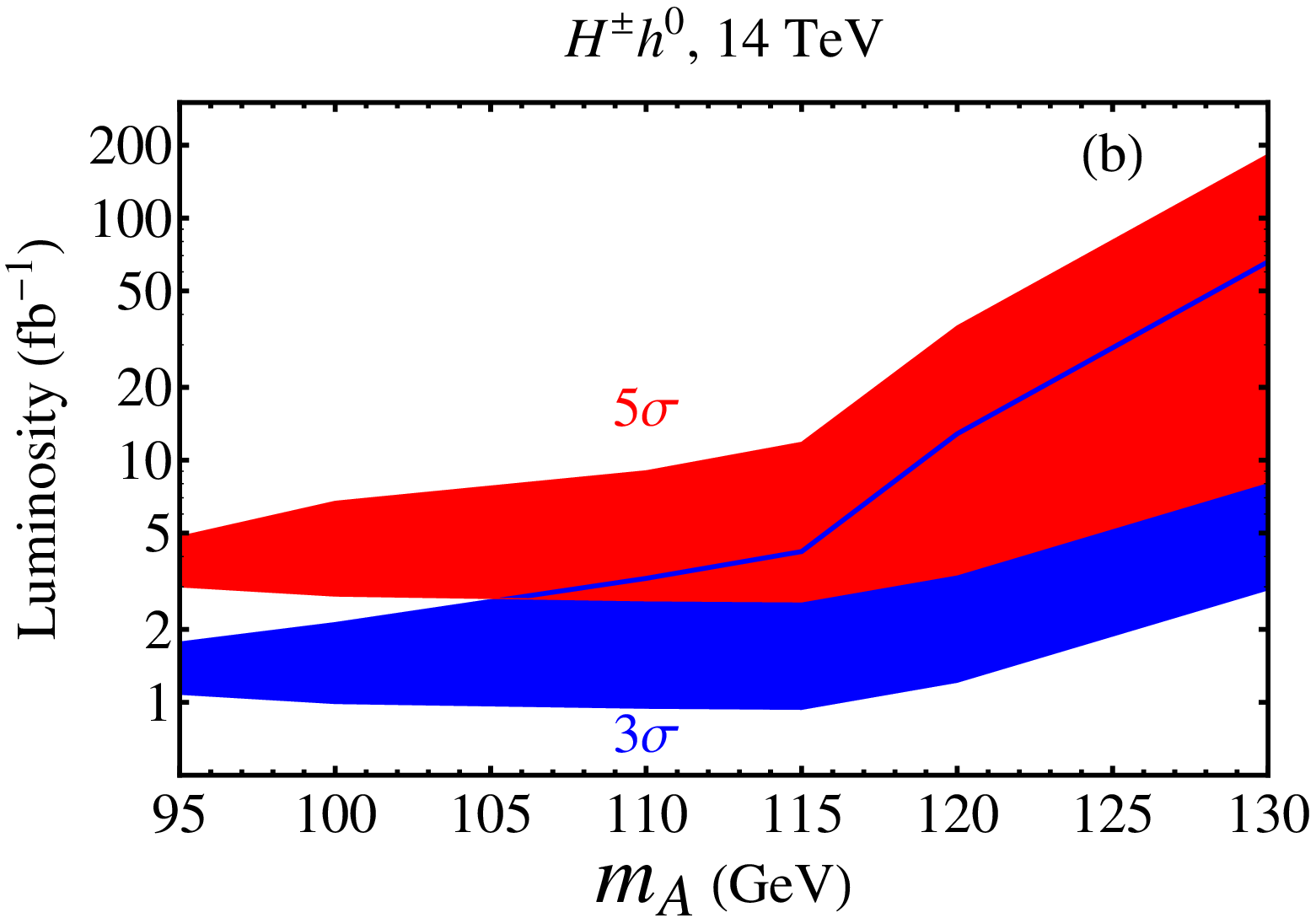}
\end{center}
\caption{Luminosity needed for $3\sigma$ (band on the bottom) and $5\sigma$ (band on the top) sensitivity as a function of $m_{A}$ for $H^\pm h^0$ at (a) 8 and (b) 14 TeV LHC.}
\label{hh+-sen}
\end{figure}

\begin{figure}[tb]
\begin{center}
\includegraphics[scale=1,width=8cm]{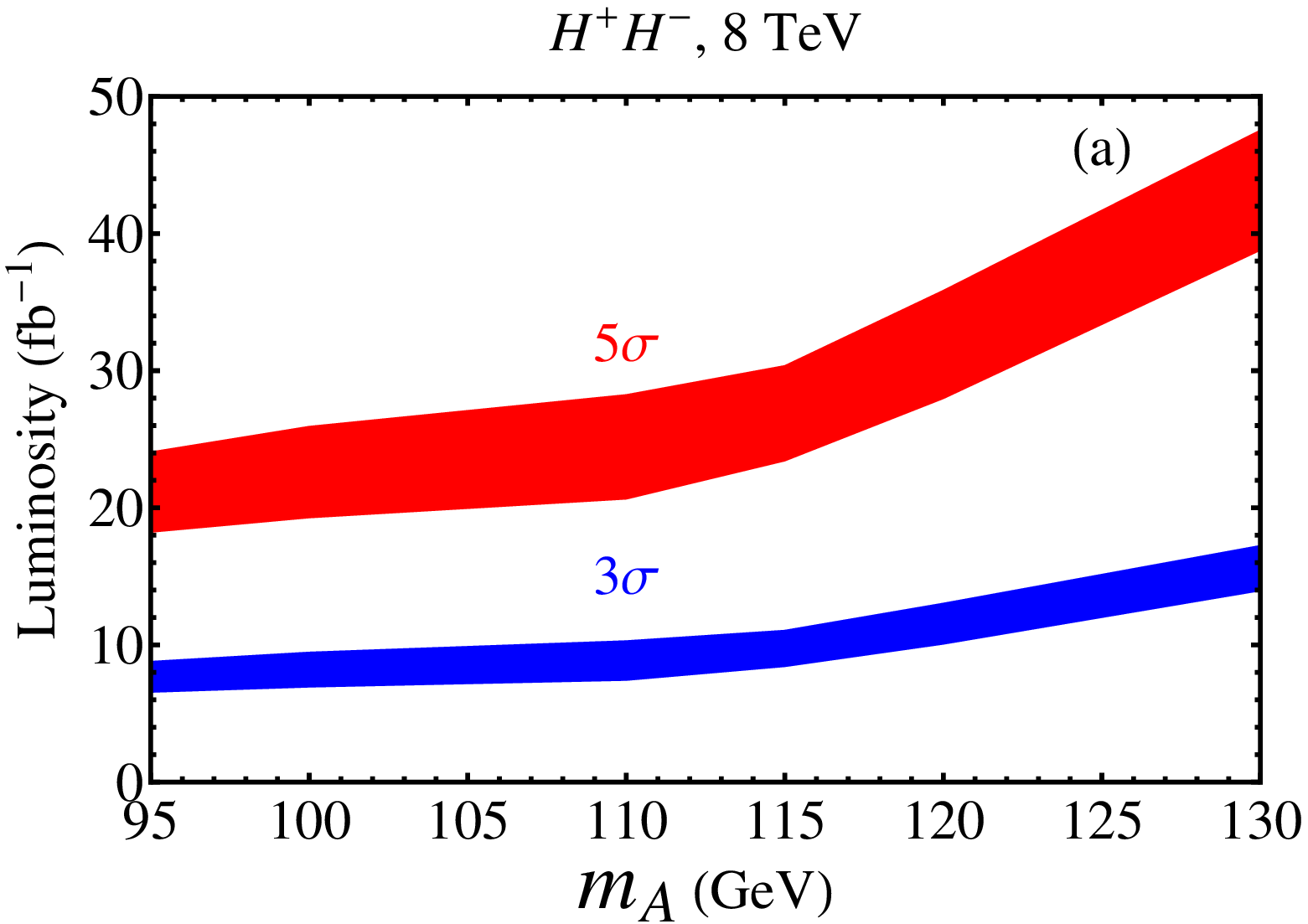}
\includegraphics[scale=1,width=8cm]{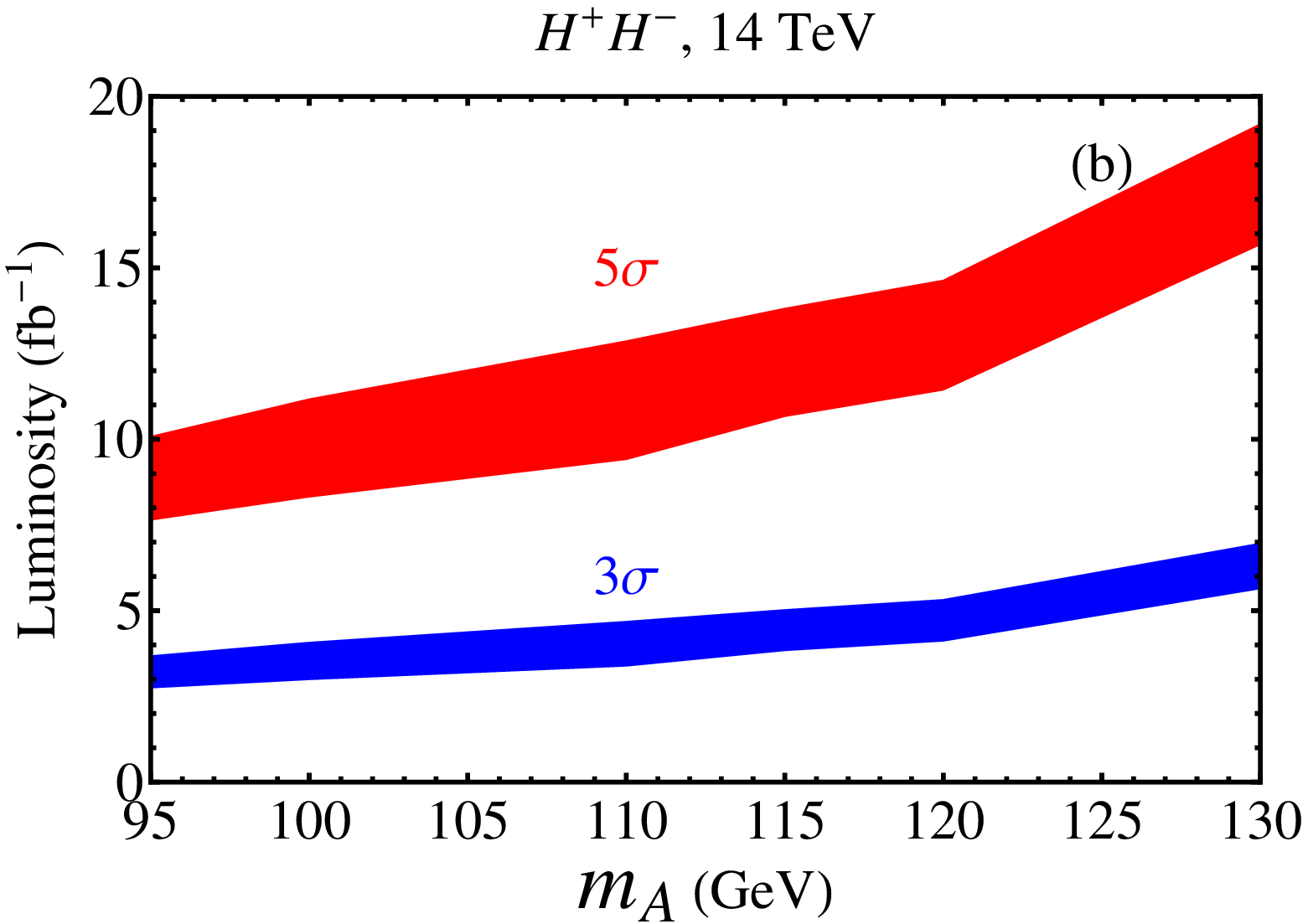}
\end{center}
\caption{Luminosity needed for $3\sigma$  (band on the bottom) and $5\sigma$ (band on the top) sensitivity as a function of $m_{A}$ for $H^+H^{-}$ at (a) 8 and (b) 14 TeV LHC.}
\label{h+h-sen}
\end{figure}

Based on the above signal and background studies, we wish to extend the exploration to a broad scope of Higgs parameter space.
We chose benchmark parameter points for a series of values of $\ma$ spanning 95 GeV to 130 GeV from our scatter points used to create Figs. \ref{totcs8-14} and \ref{BRs}.  The analysis of the previous sections was applied to these points, where Eqs. (\ref{eq:mass}) and (\ref{eq:M_tt cut}) were generalized to $\left|M_{bb}-\ma,\mh\right|<15$ GeV and $\left|M_{\tau\tau}-\ma,\mh\right|<15$ GeV, respectively.  We then estimated the span in production cross sections and branching fractions from Figs. \ref{totcs8-14} and \ref{BRs} and used those to estimate the span of integrated luminosities required for a 3$\sigma$ and 5$\sigma$ measurement of the signal after combining the pion and rho modes.  These luminosities are plotted in Figs.~\ref{ah+-sen}, \ref{hh+-sen}, and \ref{h+h-sen} as functions of $\ma$ for 8 TeV and 14 TeV.
%

\section{Summary}
\label{Sum}


Given the revolutionary discovery of the SM-like Higgs boson, it is imperative to explore the Higgs signals beyond the conventional search channels which will help in the determination of the nature of the Higgs sector.

Although the conventional signals for Higgs boson production as in Eqs.~(\ref{ggh}) and (\ref{VH}) are benefitted by leading order couplings and simple kinematics, they all depend on additional model parameters, such as
$\cos(\beta - \alpha)$.
 The Higgs boson signals from pair production $\Hpm \A$ and $H^{+}H^{-}$ as in Eq.~({\ref{gauge}), in contrast, are only governed by pure electroweak gauge interactions.
We thus consider their observability within the framework of the MSSM in the non-decoupling region.
Processes in Eq.~(\ref{mssm}) do depend on SUSY parameters, but are quite complementary to those in Eq.~(\ref{gauge}).
Since the five Higgs bosons are all relatively light in this scenario, pair production signals may be accessible.
The total cross sections for the leading pair production signal channel may range from 60 to 180 fb at 8 TeV in the mass region of our current interest, and approximately double at 14 TeV. The decay channels $\A, \h\to b\bar b$ and $\Hpm\to \tau\nu$ yield almost $100\%$ branching fractions and may provide unique final state signatures. We found that the signals for pair production are quite encouraging.

Although the SM electroweak backgrounds can be large, one of the characteristic features for the signal is the tau hadronic decays, in which the final state pions and rhos are kinematically more distinctive from the backgrounds because of the spin correlation from the decays of the charged Higgs bosons. The effects of spin correlation can be seen in the momentum distributions of Figs.~\ref{ah+-sigbkg14} and \ref{h+h-sigbkg14}. With a judicious selection of cuts, we are able to achieve quite impressive results as demonstrated in Tables \ref{ah+-events} and \ref{h+h-events}. We then generalize the analyses to a large scope of parameter space by performing the full scan over the range of Eqs.~(\ref{eq:para}) and (\ref{eq:ma}). The integrated luminosities needed to reach 3$\sigma$ and  5$\sigma$ sensitivity are shown in Figs.~\ref{ah+-sen}, \ref{hh+-sen}, and \ref{h+h-sen}.

In summary, at the 8 TeV LHC, $5\sigma$ signals for $\Hpm \A,\ \Hpm \h \to \tau^{\pm}\nu\ b\bar b$ and $H^{+}H^{-}\to \tau^{+}\nu \tau^{-}\nu$ are achievable with an integrated luminosity of 7 (11)~fb$^{-1}$ and 24 (48)~fb$^{-1}$, respectively for $\ma=95 \ (130)$ GeV. At the 14 TeV LHC, $5\sigma$ signals for
these channels would need as little as 4 (7)~fb$^{-1}$ and 10 (19)~fb$^{-1}$, respectively.
We reiterate that it is imperative to explore the Higgs signals beyond the conventional search channels which can help for the discovery and the determination of the  nature of the Higgs sector. The pair production channels $\Hpm \A$ and $H^{+}
H^{-}$ are robust processes that are independent of the model parameters.

\acknowledgments
We would like to thank Shufang Su for an early collaboration and helpful discussions.
The work of N.C.~and T.H.~was supported in part by the US Department of Energy
under grant No.~DE-FG02-12ER41832, in part by PITT PACC, and in part by the LHC-TI under U.S. National Science Foundation, grant NSF-PHY-0705682. T.L.~was supported in part by the DOE under grant No.~DE-FG02-12ER41808.



\begin{thebibliography}{99}

%

\bibitem{ATLASHnew}
ATLAS Collaboration, Phys.~Lett.~B{\bf 716}, 1 (2012).

\bibitem{CMSHnew}
CMS Collaboration, Phys.~Lett.~B{\bf 716}, 30 (2012).

%
%
%
%
%
%
%
%

\bibitem{Gunion:1989we}
  J.~F.~Gunion, H.~E.~Haber, G.~L.~Kane and S.~Dawson,
  Front.\ Phys.\  {\bf 80}, 1 (2000);
%
  J.~F.~Gunion and H.~E.~Haber,
  Nucl.\ Phys.\ B {\bf 272}, 1 (1986)
  [Erratum-ibid.\ B {\bf 402}, 567 (1993)].

\bibitem{Djouadi:2005gj}
  A.~Djouadi,
  Phys.\ Rept.\  {\bf 459}, 1 (2008)
  [hep-ph/0503173].

\bibitem{recent}
For some recent papers dealt with the MSSM aspects in light of the recent Higgs searches, see, {\it e.g.}
%
  H.~Baer, V.~Barger and A.~Mustafayev,
  arXiv:1112.3017;
  S.~Heinemeyer, O.~Stal and G.~Weiglein,
  arXiv:1112.3026;
%
  A.~Arbey, M.~Battaglia, A.~Djouadi, F.~Mahmoudi and J.~Quevillon,
  arXiv:1112.3028;
  A.~Arbey, M.~Battaglia and F.~Mahmoudi,
  arXiv:1112.3032;
  P.~Draper, P.~Meade, M.~Reece and D.~Shih,
  arXiv:1112.3068 [hep-ph];
 %
  T.~Moroi and K.~Nakayama,
  arXiv:1112.3123;
  M.~Carena, S.~Gori, N.~R.~Shah and C.~E.~M.~Wagner,
  arXiv:1112.3336 [hep-ph].
  O.~Buchmueller {\it et al.},
  arXiv:1112.3564;
  S.~Akula, B.~Altunkaynak, D.~Feldman, P.~Nath and G.~Peim,
  arXiv:1112.3645;
 %
  M.~Kadastik, K.~Kannike, A.~Racioppi and M.~Raidal,
  arXiv:1112.3647;
  J.~Cao, Z.~Heng, D.~Li and J.~M.~Yang,
  arXiv:1112.4391 [hep-ph];
%
  H.~Baer, V.~Barger and A.~Mustafayev,
  arXiv:1202.4038 [hep-ph];
  N.~Desai, B.~Mukhopadhyaya and S.~Niyogi,
  arXiv:1202.5190 [hep-ph];
%
  J.~Cao, Z.~Heng, J.~M.~Yang, Y.~Zhang and J.~Zhu,
  arXiv:1202.5821 [hep-ph].
%
%
  F.~Brummer, S.~Kraml and S.~Kulkarni,
  arXiv:1204.5977 [hep-ph];
  M.~Carena, S.~Gori, N.~R.~Shah, C.~E.~M.~Wagner and L.~-T.~Wang,
  arXiv:1205.5842 [hep-ph];
  A.~Fowlie, M.~Kazana, K.~Kowalska, S.~Munir, L.~Roszkowski, E.~M.~Sessolo, S.~Trojanowski and Y.~-L.~S.~Tsai,
  arXiv:1206.0264 [hep-ph];
  K.~Hagiwara, J.~S.~Lee and J.~Nakamura,
  arXiv:1207.0802 [hep-ph].
%

\bibitem{Christensen:2012ei}
  N.~D.~Christensen, T.~Han and S.~Su,
Phys.~Rev.~D{\bf 85}, 115018 (2012) [arXiv:1203.3207 [hep-ph]].



\bibitem{HHaber}
  H.~E.~Haber,
  hep-ph/9505240.

\bibitem{Boos:2002ze}
This is similar to the so-called ``Intense coupling region'',  except there a large $\tan\beta$ was assumed in addition:
E.~Boos, A.~Djouadi, M.~Muhlleitner and A.~Vologdin,
  Phys.\ Rev.\ D {\bf 66}, 055004 (2002)
  [hep-ph/0205160];
  E.~Boos, A.~Djouadi and A.~Nikitenko,
  Phys.\ Lett.\ B {\bf 578}, 384 (2004)
  [hep-ph/0307079].

\bibitem{Gerard:2007kn}
  J.-M.~Gerard and M.~Herquet,
  Phys.\ Rev.\ Lett.\  {\bf 98}, 251802 (2007)  [hep-ph/0703051 [HEP-PH]];
  E.~Cervero and J.~-M.~Gerard,
  arXiv:1202.1973 [hep-ph].

\bibitem{Carena:1995wu}
  M.~S.~Carena, M.~Quiros and C.~E.~M.~Wagner,
  Nucl.\ Phys.\  B {\bf 461}, 407 (1996)
  [arXiv:hep-ph/9508343];
%
  M.~S.~Carena, J.~R.~Espinosa, M.~Quiros and C.~E.~M.~Wagner,
  Phys.\ Lett.\  B {\bf 355}, 209 (1995)
  [arXiv:hep-ph/9504316].


\bibitem{Degrassi:2002fi}
  G.~Degrassi, S.~Heinemeyer, W.~Hollik, P.~Slavich and G.~Weiglein,
  Eur.\ Phys.\ J.\ C {\bf 28}, 133 (2003)
  [hep-ph/0212020];
%
  S.~Heinemeyer, W.~Hollik and G.~Weiglein,
  Eur.\ Phys.\ J.\ C {\bf 9}, 343 (1999)
  [hep-ph/9812472];
%
  M.~Frank, T.~Hahn, S.~Heinemeyer, W.~Hollik, H.~Rzehak and G.~Weiglein,
  JHEP {\bf 0702}, 047 (2007)  [hep-ph/0611326], and references therein;
%
  S.~Heinemeyer, W.~Hollik and G.~Weiglein,
  Comput.\ Phys.\ Commun.\  {\bf 124}, 76 (2000)  [hep-ph/9812320], and references therein.

\bibitem{Belyaev:2012qa}
  A.~Belyaev, N.~D.~Christensen and A.~Pukhov,
  arXiv:1207.6082 [hep-ph].


\bibitem{Bechtle:2008jh}
  P.~Bechtle, O.~Brein, S.~Heinemeyer, G.~Weiglein and K.~E.~Williams,
  Comput.\ Phys.\ Commun.\  {\bf 181}, 138 (2010)
  [arXiv:0811.4169 [hep-ph]], and references therein;
%
  P.~Bechtle, O.~Brein, S.~Heinemeyer, G.~Weiglein and K.~E.~Williams,
  Comput.\ Phys.\ Commun.\  {\bf 182}, 2605 (2011)
  [arXiv:1102.1898 [hep-ph]], and references therein.

  \bibitem{LEP2H}
Combined results from the LEP2 experiments, Phys.~Lett.~B{\bf 565}, 61 (2003).


\bibitem{CDFD0}
  T.~Aaltonen {\it et al.}  [CDF Collaboration],
  Phys.\ Rev.\ Lett.\  {\bf 103}, 101803 (2009)
  [arXiv:0907.1269 [hep-ex]];
%
  V.~M.~Abazov {\it et al.}  [D0 Collaboration],
  Phys.\ Lett.\ B {\bf 682}, 278 (2009)
  [arXiv:0908.1811 [hep-ex]].

\bibitem{ATLASH}
ATLAS Collaboration, {\tt arXiv:1202.1408};
CMS Collaboration, {\tt arXiv:1202.1488};
%
  CMS Collaboration,
  CMS-PAS-HIG-11-009;
%
  G.~Aad {\it et al.}  [ATLAS Collaboration],
  Phys.\ Lett.\  B {\bf 705}, 174 (2011)
  [arXiv:1107.5003 [hep-ex]];
%
  CMS Collaboration on $A^{0}\to \tautau$,  CMS CMS-PAS-HIG-11-029 (2011);
%
CMS Collaboration on $t\to H^{\pm} b$, CMS PAS HIG-11-008 (2011);
%
ATLAS Collaboration on $t\to H^{\pm} b$, ATLAS-CONF-2011-151 (2011).

\bibitem{LHCJuly4}
ATLAS Collaboration, ATLAS-CONF-2012-091 (2012);
ATLAS Collaboration, ATLAS-CONF-2012-092 (2012);
ATLAS Collaboration, ATLAS-CONF-2012-098 (2012);
CMS Collaboration, HIG-12-015-pas (2012);
CMS Collaboration, HIG-12-016-pas (2012);
CMS Collaboration, HIG-12-017-pas (2012).



\bibitem{QCD}
  S.~Dawson, S.~Dittmaier and M.~Spira,
  Phys.\ Rev.\ D {\bf 58}, 115012 (1998)
  [hep-ph/9805244];
  %
  S.~Dawson, T.~Han, W.~K.~Lai, A.~K.~Leibovich and I.~Lewis,
  arXiv:1207.4207 [hep-ph].

%
\bibitem{KH}
B.~K.~Bullock, K.~Hagiwara and A.~D.~Martin,
  Phys.\ Rev.\ Lett.\ {\bf 67}, 3055 (1991);
B.~K.~Bullock, K.~Hagiwara and A.~D.~Martin, Nucl.\ Phys.\ {\bf B395}, 499 (1993).

\bibitem{Boos:2005ca}
  E.~Boos, V.~Bunichev, M.~S.~Carena and C.~E.~M.~Wagner,
  eConf C {\bf 050318}, 0213 (2005) [hep-ph/0507100];
%
  M.~Guchait and D.~P.~Roy,
  arXiv:0808.0438 [hep-ph];
  A.~Ali, F.~Barreiro and J.~Llorente,
  Eur.\ Phys.\ J.\ C {\bf 71}, 1737 (2011)
  [arXiv:1103.1827 [hep-ph]];
CMS Collaboration, JINST 7 (2012) P01001 [arXiv:1109.6034v1 [physics.ins-det]].




\bibitem{smearing}
G.~L.~Bayatian et al. [CMS Collaboration], J.~Phys.~G {\bf 34}, 995 (2007); G.~Aad et al. [ATLAS Collaboration], arXiv: 0901.0512 [hep-ex].

\bibitem{MGME}
J.~Alwall, M.~Herquet, F.~Maltoni, O.~Mattelaer and T.~Stelzer, JHEP {\bf 1106}, 128 (2011).

\bibitem{Tauola}
S.~Jadach, Z.~Was, R.~Decker and J.~H.~Kuhn, Comput.\ Phys.\ Commun.\ {\bf 76}, 361 (1993).

\bibitem{NLObbw}
C.~Oleari and L.~Reina,
   JHEP {\bf 1108}, 061 (2011)
   [Erratum-ibid.\  {\bf 1111}, 040 (2011)]
   [arXiv:1105.4488 [hep-ph]].

\bibitem{NLObt}
T.~Stelzer, Z.~Sullivan and S.~Willenbrock,
   Phys.\ Rev.\ D {\bf 56}, 5919 (1997)
   [hep-ph/9705398].

\bibitem{NLOtt}
N.~Kidonakis and R.~Vogt,
   Phys.\ Rev.\ D {\bf 78}, 074005 (2008)
   [arXiv:0805.3844 [hep-ph]].

\bibitem{CP}
Q.~H.~Cao, S.~Kanemura and C.-P.~Yuan, Phys.\ Rev.\ {\bf D69}, 075008 (2004).


\bibitem{Dixon:1999di}
L.~J.~Dixon, Z.~Kunszt and A.~Signer,
  Nucl.\ Phys.\ B {\bf 531}, 3 (1998)
  [hep-ph/9803250];
  L.~J.~Dixon, Z.~Kunszt and A.~Signer,
  Phys.\ Rev.\ D {\bf 60}, 114037 (1999)
  [hep-ph/9907305].

\bibitem{Frederix:2011qg}
  R.~Frederix, S.~Frixione, V.~Hirschi, F.~Maltoni, R.~Pittau and P.~Torrielli,
  JHEP {\bf 1109}, 061 (2011)
  [arXiv:1106.6019 [hep-ph]].

\end{thebibliography}
\end{document}